\documentclass[graybox]{svmult}

\usepackage{mathptmx}       
\usepackage{helvet}         
\usepackage{courier}        
\usepackage{type1cm}        
\usepackage{makeidx}         
\usepackage{graphicx}        
\usepackage{multicol}        
\usepackage[bottom]{footmisc}

\usepackage{natbib}

\makeindex             

\begin{document}

\title*{Observing Supermassive Black Holes across cosmic time: from phenomenology to physics}
\titlerunning{Observing SMBH across cosmic time} 
\author{Andrea Merloni}
\institute{Andrea Merloni \at Max-Planck Institute of Extraterrestrial Physics, Giessenbachstr.1, 85748, Garching, Germany, \email{am@mpe.mpg.de}}
%
%
\maketitle

\abstract{In the last decade, a combination of high sensitivity, high spatial resolution observations and of
coordinated multi-wavelength surveys has revolutionized our view of extra-galactic black hole (BH) astrophysics. We now
know that supermassive black holes reside in the nuclei of almost every galaxy, grow over
cosmological times by accreting matter, interact and merge with each other, and in the process liberate
enormous amounts of energy that influence dramatically the evolution of the surrounding gas and stars,
providing a powerful self-regulatory mechanism for galaxy formation.
The different energetic phenomena associated to growing black holes and Active Galactic Nuclei (AGN), 
their cosmological evolution and the observational techniques used to unveil them, are the subject of this
chapter. In particular, I will focus my attention on the connection between the theory of high-energy astrophysical
processes giving rise to the observed emission in AGN, the observable imprints they leave at different wavelengths, 
and the methods used to uncover them in a statistically robust way. I will show
how such a combined effort of theorists and observers have led us to unveil most of the SMBH growth over a large fraction
of the age of the Universe, but that nagging uncertainties remain, preventing us from fully understating the exact role of
black holes in the complex process of galaxy and large-scale structure formation, assembly and evolution.}

\section{Introduction}
\label{sec:1}
Astrophysical black holes in the local Universe have been 
inferred to reside in two main classes of systems: X-ray binaries and active galactic nuclei (AGN). 
Gathering estimates of their masses (either directly via dynamical measurements, or indirectly, using 
phenomenological relations) allows their mass function to be derived \citep{Shankar:04,Ozel:10,Kormendy:13}.
This appears clearly bi-modal, lacking any evidence of a substantial black hole population 
at intermediate masses (i.e. between $\approx 10^2$ and $10^5 M_{\odot}$). 
While the height, width and exact mass scale of
the low-mass peak can be modeled theoretically as the end-product of
{\em stellar} (and binary) evolution, and of the physical processes that
make supernovae and gamma-ray bursts explode, the {\em supermassive}
black hole peak in this distribution is the outcome of the
cosmological growth of structures and of the evolution of mass inflow towards (and within) the 
nuclear regions of galaxies, likely modulated by the mergers these nuclear
black holes will experience as a result of the hierarchical
galaxy-galaxy coalescences.

The main question we are interested in here is the following:
given the observed population of supermassive black holes in galactic nuclei, can we constrain models
of their cosmological evolution to trace back this local population to their formation mechanisms and 
the main observable phases of growth, as identified by the entire AGN population?

As opposed to the case of galaxies, where the direct relationship between the evolving mass 
functions of the various galaxy types and the star formation distribution is not straightforward 
due to their never-ending morphological and photometric transformation, 
the case of SMBH is much simpler. By their very nature, black holes are simple ('hairless') objects, 
characterized only by two physical properties (mass and spin), the evolution of which is regulated by 
analytical formulae, to the first order functions of the rate of mass accretion onto them.
Thus, for any given ``seed'' black hole population, their full cosmological evolution can be reconstructed,
and its end-point directly compared to any local observation, 
provided that their growth phases are fully sampled observationally.

This motivates ever more complete AGN searches (surveys). The level to which the desired completeness
can in practice be achieved depends on the level 
of our understanding of the physical and electromagnetic processes that
take place around accreting black holes. So, we cannot discuss the evolution of supermassive black 
holes without an in-depth understanding of AGN surveys, and of their results; but at the same time, 
we cannot understand properly these surveys if we do not understand the physics behind the observed
AGN phenomenology.  

In this chapter, I review the current state of affairs regarding the study of the evolution of the black
hole population in the nuclei of galaxies. I will first (\S~\ref{sec:surveys_agn}) describe the observational techniques used to survey
the sky in search of signs of black hole activity, and the progresses made on constraining the phenomenological
appearance of AGN (\S~\ref{sec:obs_agn}). Then, in section~\ref{sec:sed_phys}, I will move from the phenomenological to the physical 
description of the processes thought to be responsible for the observed Spectral Energy Distribution (SED) in luminous AGN, 
focusing in particular 
on the properties of AGN accretion discs (\S~\ref{sec:adisc}), coronae (\S~\ref{sec:coronae}), and the IR-emitting dusty
obscurer (the so-called ``torus'', \S~\ref{sec:torus}). 
In section~\ref{sec:lf} I will present a concise overview of the current
state of the art of AGN luminosity function studies at various wavelength, encapsulating our knowledge about the overall population
cosmic evolution. The final section (\S~\ref{sec:soltan}) is devoted to a general discussion of the so-called {\em Soltan argument},
i.e. the method by which we use the evolutionary study of the AGN population to infer additional global physical properties of the 
process of accretion onto and energy release by supermassive black holes. In particular, I will show how robust limits on the average
radiative and kinetic efficiency of such processes can be derived.

A few remarks about this review are in place. First of all, I do not discuss in detail here the physics of relativistic jets in AGN, 
which carry a negligible fraction of the bolometric output of the accretion process, but can still carry large fractions of the
energy released by the accretion process in kinetic form. This is of course a
complex and rich subject in itself, and I refer the reader to the recent reviews 
by \citet{ghisellini:11}, \citet{Perlman:13} and \citet{Heinz:14}. Nonetheless, I will include a discussion
about radio luminosity function evolution, which is functional to the aim of compiling a census of the kinetic energy output of SMBH
over cosmic time. I also do not discuss in any detail the impact growing black holes might have on the larger-scale systems they are
embedded in. The generic topic of {\em AGN feedback} has been covered by many recent reviews 
\citep[see e.g.][]{cattaneo:09,fabian:12,mcnamara:12,merloni:13}, and would definitely deserve more space than is allowed here. 
Finally, part of the material presented here has been published, in different form, in two recent reviews \citep{merloni:13,gilfanov:14}, 
and in \citet{merloni:14}.

\section{Finding supermassive black holes: surveys, biases, demographics}
\label{sec:surveys_agn}

Accretion onto supermassive black holes at the center of galaxies manifests itself in a wide
variety of different phenomena, collectively termed Active Galactic
Nuclei. Their luminosity can reach values orders of magnitude larger than the collective radiative
output of all stars in a galaxy, as in the case of powerful Quasars (QSO), reaching the Eddington luminosity for
black holes of a few billion solar masses\footnote{The Eddington luminosity 
is defined as $L_{\rm Edd}=4 \pi G M_{\rm BH} m_{\rm p} c /
\sigma_{\rm T} \simeq 1.3 \times 10^{38} (M_{\rm BH}/M_{\odot})$ ergs
s$^{-1}$, where $G$ is the Newton constant, $m_{\rm p}$ is the proton mass, $c$ the speed of light 
and $\sigma_{\rm T}$ the Thomson scattering cross section.}, which can be visible at the highest 
redshift explored ($z>7$). On the other hand, massive black holes in galactic nuclei can be 
exceedingly faint, like in the case of Sgr A* in the nucleus of the Milky Way, which radiates at less than a 
billionth of the Eddington luminosity of the $6.4 \times 10^6$ M$_{\odot}$ BH harboured there.
Such a wide range in both black hole masses and accretion rates of SMBH results in a wide, 
complex, observational phenomenology. 

The observational characterization of the various accretion 
components is challenging, because of the
uniquely complex multi-scale nature of the problem.
Such a complexity greatly affects our ability
to extract reliable information on the nature of the accretion
processes in AGN and does often introduce severe observational biases,
that need to be accounted for when trying to recover the underlying
physics from observations at various wavelengths, either of individual
objects or of large samples. 

Simple order-of-magnitudes evaluations will suffice here. 
Like any accreting black holes, an AGN releases most of
its energy (radiative or kinetic) on the scale of a few Schwarzschild
radii ($\sim 10^{-5}$ pc for a $10^8$ $M_{\odot}$ BH). However, 
the mass inflow rate (accretion rate) is determined by the galaxy ISM properties at
the location where the gravitational influence of the central
black hole starts dominating the dynamics of the intergalactic gas (the so-called Bondi radius), 
some $10^5$ times further
out. The broad permitted atomic emission lines that are so prominent in the optical
spectra of un-obscured QSOs 
are produced at $\sim$0.1-1 pc (Broad Line Region, BLR), while, 
on the parsec scale, and
outside the sublimation radius, a dusty, large-scaleheight, possibly
clumpy, medium obscures the view of the inner engine \citep{elitzur:08}
crucially determining the observational properties of the AGN
\citep{netzer:08}; on the same scale, powerful star formation might be triggered  
by the self-gravitational instability of the inflowing gas \citep{goodman:03}. Finally, 
AGN-generated outflows (either in the form of winds or relativistic
jets) are observed on galactic scales and well above (from a few
to a few hundreds kpc, some
$\sim 10^8-10^{10}$ times $r_{\rm g}$!), often carrying substantial amounts
of energy that could dramatically alter the (thermo-) dynamical state
of the inter-stellar and inter-galactic medium.

When facing the daunting task of assessing the cosmological evolution of
AGN, i.e. observing and measuring the signs of accretion onto nuclear SMBH within distant
galaxies, it is almost impossible to achieve the desired high spatial resolution throughout
the electromagnetic spectrum, and one often resorts to less direct
means of separating nuclear from galactic light.
There is, however, no simple prescription for efficiently performing
such a disentanglement: 
the very existence of scaling relations between black holes and their
host galaxies,
and the fact that, depending on the specific physical condition
of the nuclear region of a galaxy at different stages of its
evolution, the amount of matter captured within the Bondi radius
can vary enormously,
imply that growing black holes will always
display a large range of ``contrast'' with the host galaxy light.
 
\subsection{On the AGN/galaxy contrast in survey data}

More specifically, let us consider an
AGN with optical B-band luminosity given by $L_{\rm B,AGN}=\lambda L_{\rm Edd}
f_{\rm B}$, where we have introduced the Eddington ratio ($\lambda\equiv L_{\rm bol}/L_{\rm
  Edd}$), and a bolometric correction 
$f_{\rm B}\equiv L_{\rm B,AGN}/L_{\rm bol}\approx 0.1$
\citep{richards:06}.  Assuming a mean black hole to host galaxy mass ratio of
$A_0 = 0.002$, the
contrast between nuclear AGN continuum and host galaxy blue light is
given by:
\begin{equation}
\label{eq:agn_contrast}
\frac{L_{\rm B,AGN}}{L_{\rm
    B,host}}=39 \lambda \left(\frac{f_{\rm B}}{0.1}\right) \left(\frac{A_0}{0.002}\right)\frac{(M_*/L_{\rm B})_{\rm
    host}}{3(M_{\odot}/L_{\odot})}
\end{equation} 

Thus, for typical mass-to-light ratios, the AGN will become
increasingly diluted by the host stellar light in the rest-frame UV-optical bands 
at Eddington ratios
$\lambda$ smaller than a few per cent. 

Similar considerations can be applied to the IR bands, as follows.
For simplicity, we use here the \citet{Rieke:09} 
relation between monochromatic (24$\mu$m) IR luminosity and Star Formation Rate (SFR, expressed
in units of solar masses per year): $L_{\rm 24,SFR}\approx 5 \times 10^{42} \times {\rm SFR}$, and the
\citet{Whitaker:12} analytic expression for the ``main sequence'' of star forming galaxies:
$\log SFR=\alpha(z)*(\log M_*-10.5)+\beta(z)$, with $\alpha(z)=0.7-0.13z$ and $\beta(z)=0.38+1.14z-0.19z^2$.
Then, the rest-frame 24$\mu$m ``contrast'' between AGN and galaxy-wide star formation ca be written as:

\begin{equation}
\frac{L_{\rm IR, AGN}}{L_{\rm IR, SF}}\approx 160 \lambda 10^{-\beta(z)} \left(\frac{f_{24}}{0.1}\right) \left(\frac{A_0}{0.002}\right) \left(\frac{M_*}{10^{10.5}M_{\odot}}\right)^{1-\alpha(z)} \,,
\end{equation}
where we have defined $f_{24}$ the AGN bolometric correction at 24$\mu$m, and $A_0$ is here assumed, for simplicity, 
to be redshift independent.
Thus, for a ``typical'' $10^8 M_{\odot}$ black holes in a $10^{10.5} M_{\odot}$ main-sequence star-forming host,
the IR emission produced by the global star formation within the galaxy dominates over the AGN emission for $\lambda < 0.13$, or $\lambda < 0.015$, at $z\sim 1$ or $z\approx 0$, respectively.

When considering star-formation induced hard (2-10 keV) X-ray emission, instead, we obtain
\begin{equation}
\label{eq:xlim}
\frac{L_{\rm X, AGN}}{L_{\rm X, SF}}\approx 10^5 \lambda 10^{-\beta(z)} \left(\frac{f_{X}}{0.03}\right) \left(\frac{A_0}{0.002}\right) \left(\frac{M_*}{10^{10.5}M_{\odot}}\right)^{1-\alpha(z)} \,,
\end{equation}
where  $f_{X}$ the AGN bolometric correction from the 2-10 keV band, and we have used the 
expression $L_{\rm X,SF}\simeq 2.5 \times 10^{39} \times {\rm SFR}$ \citep{Ranalli:03,Gilfanov:04}: for the same level of SF in a main sequence AGN host, the nuclear AGN emission dominates the hard X-ray flux as long as the accretion rate exceeds $\lambda > 2 \times 10^{-4}$, or $\lambda > 2\times 10^{-5}$, at $z\sim 1$ or $z\approx 0$, respectively.

This has obvious implications for our understanding of selection biases in AGN surveys.
It is clear then that the most luminous optical QSOs (i.e. AGN shining at bolometric luminosity
larger than a few times 10$^{45}$ erg/s), 
represent just the simplest case, as their light out-shines the emission from the host galaxy, 
resulting in point-like emission with peculiar colors. Less
luminous, Seyfert-like, AGN will have a global SED with a
non-negligible contribution from the stellar light of the host. As a
result, unbiased AGN samples extending to lower-luminosities
will inevitably have optical-NIR colors spanning a
large range of intermediate possibilities between purely accretion-dominated 
and purely galaxy-dominated.
Optical (and, to a large extent, NIR) surveys will easily pick up AGN
at high Eddington ratio, and thus, potentially, all members of a
relatively homogeneous class of accretors, while
deep X-ray (and radio) surveys can circumvent such biases, by detecting and
identifying accretion-induced emission in objects of much lower Eddington ratio.

\begin{figure*}
 \includegraphics[width=\textwidth]{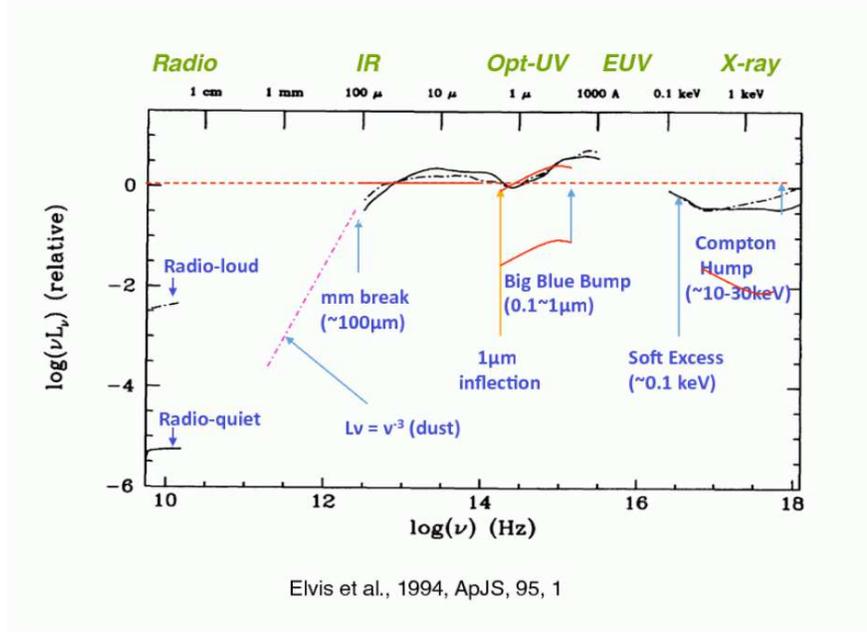}
\caption{Mean Spectral Energy distribution of UV/X-ray selected quasar
from \citet{elvis:94}. Solid black line is for radio-quiet QSOs,
dot-dashed for radio-loud.}
\label{fig:elvis}       
\end{figure*}

\subsection{Phenomenology of AGN Spectral Energy Distributions}
\label{sec:obs_agn}

The process of identifying AGN embedded within distant (or nearby) galaxies that
we have discussed above is intimately connected with the meticulous work
needed to piece together their Spectral Energy Distribution (SED) across the electromagnetic spectrum.

For the practical reasons discussed in the previous section, 
up until recent years accurate SED of accreting SMBH were
constructed mainly from bright un-obscured (type-1) 
QSO samples. Setting the standard for
almost 20 years, the work of \citet{elvis:94}, based on a
relatively small number (47) of UV/X-ray selected quasars, has been
used extensively as a template for the search and characterization of
nearby and distant AGN. The Elvis et al. (1994) SED is dominated by AGN accreting at the highest Eddington ratio, and,
as shown in Figure~\ref{fig:elvis}, this
spectral energy distribution is characterized by a relative flatness
across many decades in frequency, with superimposed two prominent
broad peaks: one in the UV part of the spectrum (the so-called Big
Blue Bump; BBB), one in the Near-IR, separated from an inflection point at
about 1$\mu$m.

Subsequent investigations based on large, optically selected QSO
samples (most importantly the SDSS one, Richards et al. 2006) have
substantially confirmed the picture emerged from the Elvis et
al. (1994) study. Apart form a difference in the mean X-ray-to-optical
ratio (optically selected samples tend to be more optically bright
than X-ray selected ones, as expected), the SDSS quasars have indeed 
a median SED similar to those shown in fig.~\ref{fig:elvis} for AGN accreting at 
$L/L_{\rm Edd}>0.1$,
despite the difference in redshift and
sample size.

\begin{figure*}
\centering
\begin{tabular}{cc}
\includegraphics[width=0.47\textwidth]{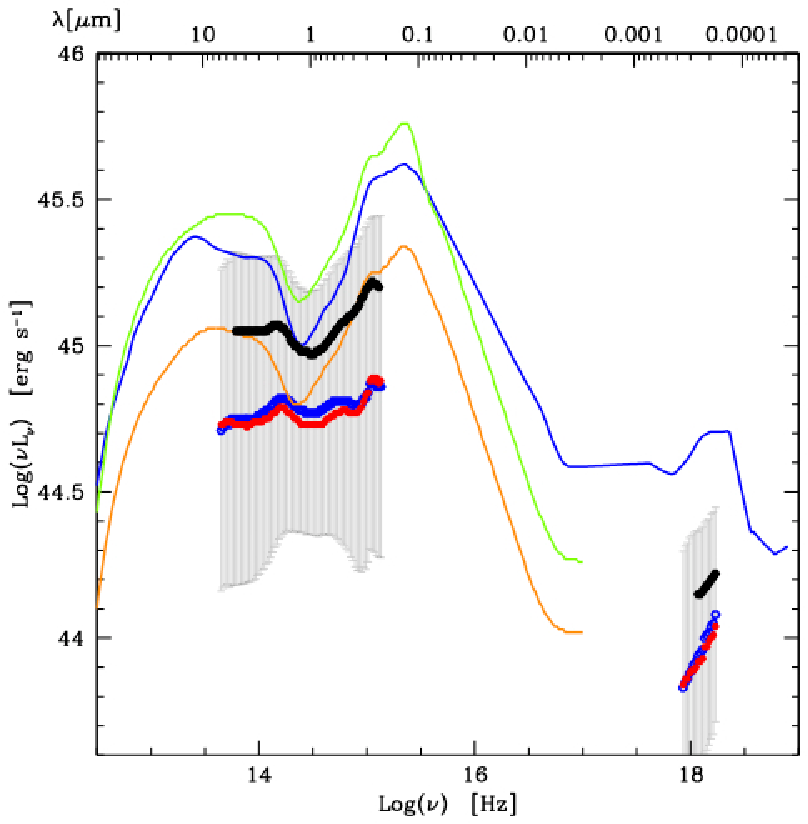}&
\includegraphics[width=0.47\textwidth]{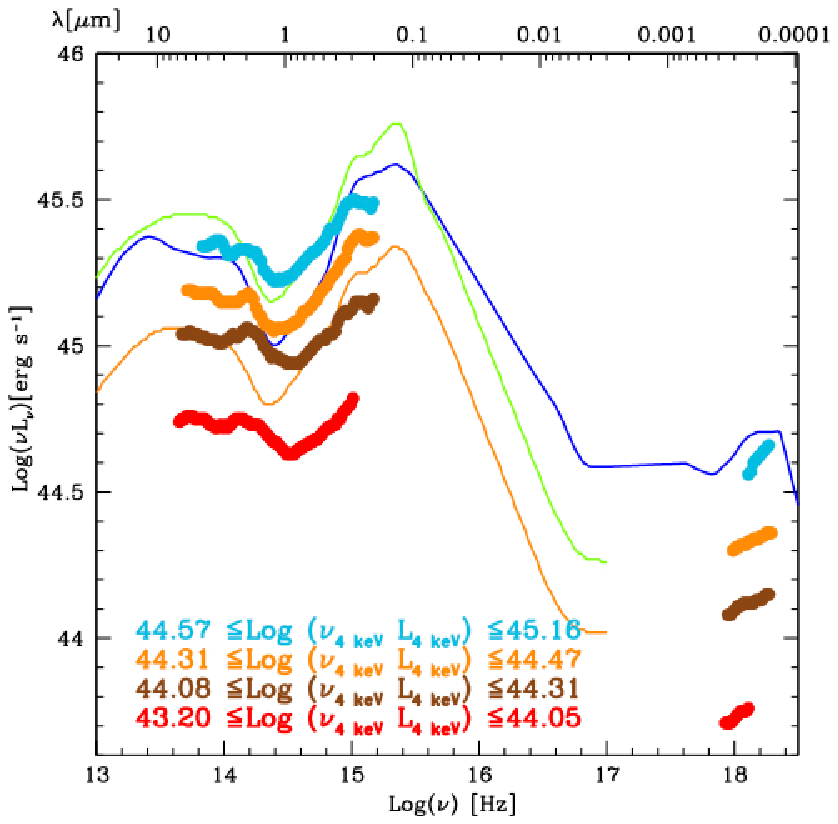}\\
\end{tabular}
\caption{{\it Left Panel:} The median (red points) and the mean (blue
  points) SEDs for the total spectroscopic type-1 COSMOS AGN sample. 
  The mean SED for
  the ``Pure'' sample (estimated galaxy contamination $<10$\%) 
  is represented with black points. The error bars
  (gray area) represent the dispersion of the total spectroscopic
  sample around the mean SED. The average SEDs are compared with the
  mean SED of Elvis et al. (1994, blue line), the mean SEDs of
  Richards et al. (2006) using all the SDSS quasar sample (green line)
  and the near-IR dim SDSS quasar sample (orange line). {\it Right
    Panel:} The median SEDs computed splitting the ``Pure'' 
  sample in bins of increasing 
  X–ray luminosity at 4 keV. From \citet{lusso_phd:11}}
\label{fig:lusso_sed}
\end{figure*}

Large multi-wavelength galaxy survey and extensive follow-up
campaigns of medium-wide and deep X-ray surveys (such as the Chandra
Deep Field, \citealt{giacconi:02}; the COSMOS field, \citealt{hasinger:07}; or
the X-Bootes field, \citealt{hickox:09}) have allowed to extend AGN SED
systematic studies to a wide variety of Eddington ratios and AGN/galaxy
relative contributions. Figure~\ref{fig:lusso_sed} shows the results of 
\citet{lusso_phd:11} 
(see also \citealt{lusso:10} and \citealt{elvis:12}), who
analysed an X-ray selected sample of AGN in the
COSMOS field, the largest fully identified and
redshift complete AGN sample to date. When restricted to a ``pure''
QSO sample (i.e. one where objects are pre-selected on the basis of a
minimal estimated galaxy contamination of $<10$\%), the SED of the
COSMOS X-ray selected AGN is reminiscent of the Elvis et al. (1994) 
and Richard et al. (2006) ones, 
albeit with a less pronounced inflection point at
1$\mu$m. The mean (and median) SED for the 
whole sample, however, apart from having a lower average
luminosity, is also characterized by much less pronounced UV and NIR
peaks. This is indeed expected whenever stellar light from the host
galaxy is mixed in with the nuclear AGN emission.

\begin{figure}
  \centering
  \resizebox{!}{0.8\textwidth}{\includegraphics{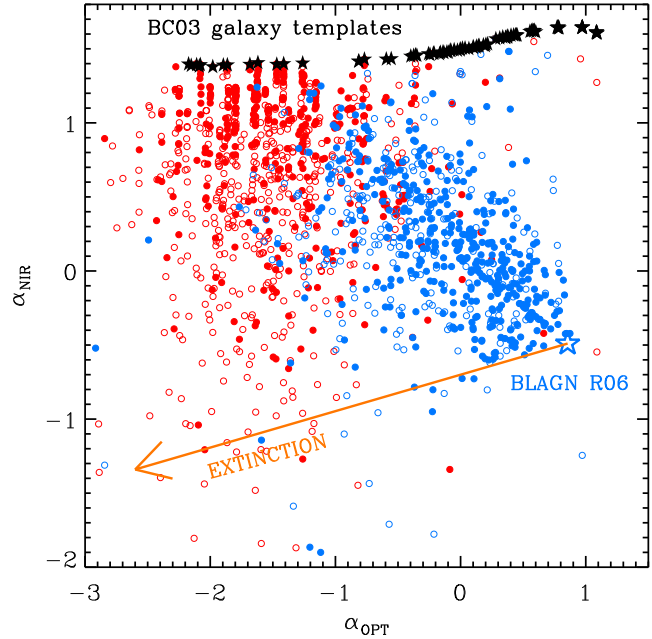}}
  \caption{Observed rest-frame SED slopes in the optical ($\alpha_{\rm
      OPT}$, between 0.3 and 1 $\mu$m) and NIR ($\alpha_{\rm NIR}$
    between 1 and 3 $\mu$m) for all ($\sim$1650) X-ray selected AGN in the
    COSMOS survey.  Blue filled circles denote spectroscopically
    confirmed type 1 (broad lined) AGN, blue empty circles denote
    candidate type 1 AGN from the photo-z sample.  Red filled circles
    are spectroscopically confirmed type 2 (narrow lined) AGN, empty
    red circles are candidate type 2 AGN from the photo-z sample.  The
    empty blue star marks the colors of a pure intrinsic type 1 quasar
    SED (from \citealt{richards:06}), while black stars are the loci
    of synthetic spectral templates of galaxies, with increasing
    levels of star formation form the left to the right.  Nuclear
    obscuration moves every pure type 1 AGN along the direction of the orange
    arrow. From \citet{bongiorno:12}.}
   \label{fig:sed_cosmos}
\end{figure}

Figure~\ref{fig:sed_cosmos} \citep{bongiorno:12,hao:12} 
further illustrates this point. 
It displays the slope of the
rest-frame SED in the optical ($\alpha_{\rm OPT}$, between 0.3 and 1
$\mu$m) and NIR ($\alpha_{\rm NIR}$ between 1 and 3 $\mu$m) bands, i.e. long- and short-wards of the 
$\sim 1\mu$m inflection point. Pure
QSOs, i.e., objects in which the overall SED is dominated by the
nuclear (AGN) emission would
lie close to the empty blue star in the lower right corner (positive
optical slope and negative NIR slope).  The location of the X-ray
selected AGN in Figure~\ref{fig:sed_cosmos} clearly shows instead that, in
order to describe the bulk of the population, one needs to consider
both the effects of obscuration (moving each pure QSO in the direction
of the orange arrow) and an increasing contribution from galactic stellar
light (moving the objects towards the black stars in the upper
part of the diagram).

\section{The spectral components of AGN: accretion discs, coronae and dusty tori}
\label{sec:sed_phys}

In the previous section, we have presented a phenomenological view of AGN Spectral Energy Distributions,
as can be gained by multi-wavelength AGN/QSO surveys, without paying too much attention to the physical origin
of the main spectral components themselves.
In this section, instead, we analyse the main spectral components of AGN SED, to highlight the connections
between AGN phenomenology and physical models of accretion flows.

\subsection{AGN accretion discs}
\label{sec:adisc}
The gravitational energy of matter dissipated in the accretion flow around 
a black hole  is primarily converted to photons of UV and soft X-ray
wavelengths. 
The lower limit on the characteristic temperature  of the emerging  radiation can be 
estimated assuming the most
radiatively efficient configuration: an optically thick accretion
flow. Taking into account that the size of the emitting region is
$r\sim 10r_{\rm g}$ ($r_{\rm g}=GM_{\rm BH}/c^2$ is the gravitational radius) and assuming a black body
emission spectrum one obtains: 
\begin{eqnarray}
kT_{\rm bb}= \left(\frac{L_{\rm bol}}{\sigma_{\rm SB}\pi r^2}\right)^{1/4}
\approx 14 \left(\frac{L_{\rm bol}}{10^{44}}\right)^{1/4}     \left(\frac{M_{\rm BH}}{10^8}\right)^{-1/2}  {\rm eV} 
\end{eqnarray}

Proper treatment of the angular momentum transport within the accretion flow allows a full analytical
solution of optically thick (but geometrically thin) discs, first discovered by Shakura \& Sunyaev (1973).
It is a major success of their theory the fact that, for typical AGN masses and
luminosities (and thus accretion rates), the expected spectrum of the
accretion disc should peak in the optical-UV bands, 
as observed. Indeed, a
primary goal of AGN astrophysics in the last decades has been to
model accurately the observed shape of the BBB in terms of standard
accretion disc models, and variations thereof.

The task is complicated by at least three main factors. 
First of all,
standard accretion disc theory, as formulated by \cite{shakura:73},
 needs to be supplemented by a description of the disc vertical structure and, in
particular, of its atmosphere, in order to accurately predict
spectra. This, in turn, depends on the exact nature of viscosity and
on the micro-physics of turbulence dissipation within the
disc. As in the case of XRB, models for geometrically thin and optically thick
AGN accretion discs has been calculated to increasing levels of
details, from the simple local blackbody approximation to stellar 
atmosphere-like models where the vertical structure and the local
spectrum are calculated accounting for the major radiative transfer
processes (e.g. the {\tt TLUSTY} code of \citealt{hubeny:00}).

\begin{figure*}
\centering
\includegraphics[width=0.8\textwidth]{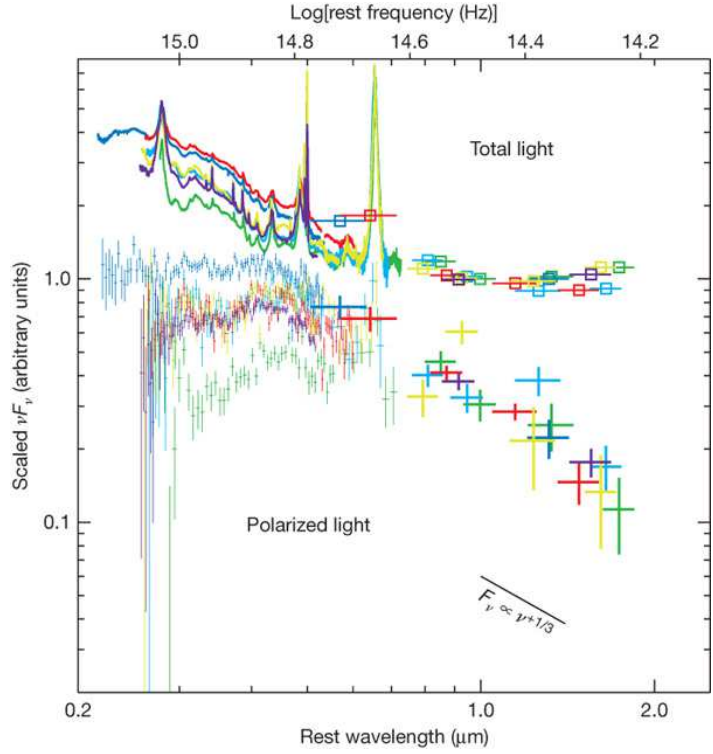}\\
\caption{Total-light spectra of nearby QSOs, shown as bold traces in the optical and as squares in the near-infrared,
 and normalized at 1 $\mu$m in the rest frame. 
Polarized-light spectra (arbitrarily shifted by a factor of 3 with respect to the total light, for clarity), 
are shown as light points in the optical and as bold points in the near-infrared, 
also separately normalized at 1 $\mu$m, by fitting a power law to the near-infrared polarized-light spectra. 
For both total-light and polarized-light data, horizontal bar lengths indicate bandwidth. 
The total-light spectra begin to increase in $\nu F_{\nu}$ at wavelengths around, 
or slightly greater than, 1 $\mu$m. In contrast, the polarized-light spectra all consistently and 
systematically decrease towards long wavelengths, showing a blue shape of approximately power-law form. 
From \citet{Kishimoto:08}}
\label{fig:kishimoto}
\end{figure*}

\begin{figure*}
\centering
\includegraphics[width=0.7\textwidth]{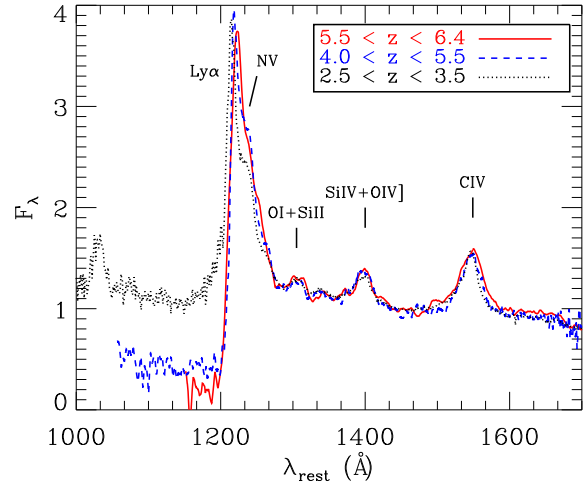}\\
\caption{Stacked spectra
  of quasars in different redshift bins.  Note that the relative
  intensity of the metal lines (and in particular the
  (SiIV+OIV])$/$CIV ratio) remains constant over the wide redshift
  interval 2.5$<z<$6.4, indicating that the metallicity in the observed
  quasars does not evolve with redshift.  From \cite{juarez:09}.}
\label{fig:fan08}
\end{figure*}

A second complicating effect, a purely observational one, is the fact
that the intrinsic disc continuum emission is often buried underneath
a plethora of permitted atomic emission lines, 
many of which broadened significantly 
by gas motions in the vicinity of the central black hole (see, e.g. the solid lines of Figure~\ref{fig:kishimoto}). 
Interestingly, the metallicities implied by the
relative strength of broad emission lines do not show any
significant redshift evolution (see Fig.~\ref{fig:fan08}): 
they are solar or super-solar, even in
the highest redshift QSOs known \citep[see e.g.][]{hamann:92}, in contrast with the
strong evolution of the metallicity in star forming galaxies.

In particularly
favorable geometrical observing conditions, however, by looking at 
optical spectra in polarised light the ``contaminating'' broad emission lines
are removed, and the true continuum of the accretion disc is revealed. This shows a broad dip
possibly corresponding to the Balmer edge absorption expected from an
accretion disc atmosphere \citep{Kishimoto:03}. 
Extending the polarised continuum into the near-IR reveals the classic
long wavelength $\nu^{1/3}$ spectrum expected from simple accretion disc
models \citep[see][and figure~\ref{fig:kishimoto}]{Kishimoto:08}.

Finally, a third complicating factor is that the real physical condition in the inner few
hundreds of Schwarzschild radii of an AGN might be more
complex than postulated in the standard accretion disc model: for example,
density inhomogeneities resulting in  cold, thick clouds 
 which reprocess the intrinsic continuum have been considered at
various stages as responsible for a number of observed mismatches
between the simplest theory and the observations 
\citep[see e.g.][and references therein]{guilbert:88,merloni:06,lawrence:12}.

Essentially all of the above mentioned problems are particularly severe in the UV part of
the spectrum, where observations are most challenging. 
\citep{shang:05} compared broad-band UV-optical accretion
disc spectra from observed quasars with accretion disc models.
They compiled quasi-simultaneous QSO spectra 
in the rest-frame energy range 900-9000 $\AA$ and fitted their continuum 
emission with broken power-law models, and then compared the behavior 
of the sample to those of non-LTE thin-disk models 
covering a range in black hole mass, Eddington ratio, disk inclination, 
and other parameters. The results are far from conclusive: on the one hand,
the observed slopes are in general consistent with the expectations of
sophisticated accretion disc models. On the other hand, the
spectral UV break appears to always be around 1100 $\AA$, and does not
scale with the black hole mass in the way expected.

\begin{figure}
\begin{center}
\includegraphics[width=0.8\textwidth]{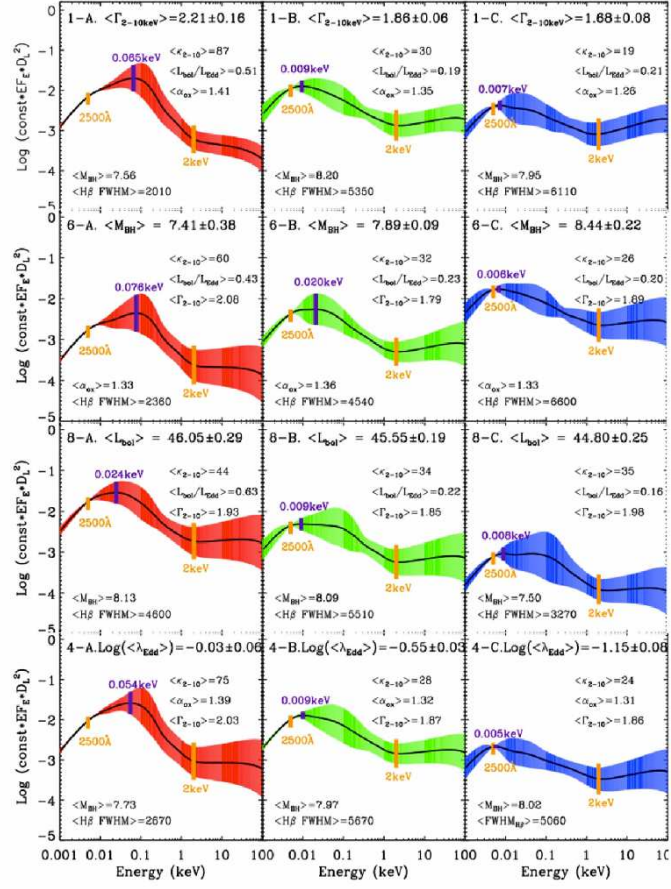}
\end{center}
\caption{The AGN mean SEDs based on different values of 4 parameters from 
SED model fitting of AGN (from top to bottom:
2-10 keV X-ray spectral index, black hole mass, bolometric
luminosity and Eddington ratio. For each parameter, 
the 51 sources are sorted according to the parameter value, and  re-normalized
 SED to the mean luminosity at 2500 $\AA$ of. The three panels (A, B, C) in 
each row show the mean SEDs for the subsets classified by the parameter 
shown in the panel title. In each panel the solid curve is the mean SED, 
while the shaded coloured region is the 1$\sigma$ deviation. 
The peak position of the SED is marked by the vertical solid purple line.
 The average values of some other parameters in that subset are also shown
 in the panel. From Jin et al. (2012)}
\label{fig:jin}
\end{figure}

\cite{jin:12}, on the other hand, have looked at detailed continuum fits 
to the joint optical-UV-X-ray SED of 51 nearby AGN with known black hole 
masses. Figure ~\ref{fig:jin} shows the variation of the mean SED as a
function of various parameters, such as X-ray spectral index $\Gamma_{\rm 2-10}$, black hole mass, bolometric luminosity and Eddington ratio.
Clearly, global trends in the basic disc properties 
(such as its peak temperature) are observed, in correlation with the main 
parameter of the accretion flow. From a Principal Component analysis,
Jin et al (2012) found that the first two eigenvectors contain
$\sim 80$\% of all correlations in the matrix, with the first one 
strongly correlating with black hole mass, and the second one with the bolometric 
luminosity, while both correlate with the Eddington ratio. Interestingly,
this turns out to be consistent with the results of a PCA analysis of 
the emission line-dominated QSO spectra \citep{boroson:02}.

Having a well-sampled spectral energy distribution for the AGN emission produced by a standard,
Shakura \& Sunyaev (1973) accretion disc around a SMBH of known mass, 
could in principle lead to useful constraints
on the overall radiative efficiency of the accretion process, and therefore on the nature 
of the inner boundary condition of the accretion disc, and on the black hole spin itself.

\cite{davis:11} have made a first systematic attempt to estimate the radiative 
efficiencies of accretion discs in a sample of QSOs. In individual AGN, 
thin accretion disk model spectral fits can be used to infer the total rate of mass accretion  
onto the black hole $\dot M$, if its mass $M_{\rm BH}$ is known. In fact, by measuring the 
continuum disk luminosity in the optical band (i.e. in the Rayleigh-Jeans part of the 
optically thick multi-color disc spectrum), the accretion rate estimates are relatively 
insensitive to the actual model of the disc atmosphere.
The principle is analogous to that employed in black hole X-ray binaries in order to constrain BH spin from the
disc continuum measurements \citep[see][]{Mcclintock:14}, but in a typical AGN, 
the above-mentioned observational intricacies need to be dealt with, together with the fact that
the BBB is much less well sampled than in a stellar mass black hole. On the other hand, the 
uncertainty in the distance to the object, that plagues the studies of galactic black holes
is not an issue for QSOs with measured spectroscopic redshift. Very massive black holes at high redshift
have the peak of the disc emission well in the optical bands. Provided one is able to properly correct for the increasing
optical depth of the Inter-galactic medium, it could be possible to use simple photometric SED modelling
to constrain properties of the disc and the central black hole (see
e.g. \citealt{Ghisellini:10}). At the opposite end of the mass spectrum, 
small mass black holes accreting at very high rate in the local universe
are expected to have such a high disc temperature that the tail of the optically thick thermal emission should appear 
as ``soft excess'' in the soft X-ray energy bands. \cite{Done:13} have indeed argued that some Narrow-Line Seyfert 1 galaxies 
can indeed be modelled with a very high temperature accretion disc, thus gaining information on the disc inner
 boundary, and, indirectly,
on the BH spin.

\subsection{AGN coronae and X-ray spectral properties}
\label{sec:coronae}

The upper end of the relevant temperature range reached by accretion flows onto black holes
is achieved in the limit of optically thin emission from a hot plasma, 
possibly analogous to the solar corona, hence the name of accretion disc ``coronae'' \citep{galeev:79}. 
The virial temperature of particles near a black hole,
$kT_{\rm vir}=GM_{\rm BH} m/r\propto mc^2/(r/r_{\rm g})$, does
not depend on the black hole mass, but only on the mass
of the particle $m$, being $T_{\rm vir,e}\sim 25(r/10r_{\rm g})^{-1}$
keV for electrons and $T_{\rm vir,p}\sim 46(r/10r_{\rm g})^{-1}$
MeV for protons. As the electrons are the main radiators that determine
the emerging spectral energy distribution, while the protons (and ions) are the 
main energy reservoir, the outcoming radiation temperature for optically thin flows
depends sensitively on the detailed micro-physical mechanisms through which 
ions and electrons exchange their energy in the hot plasma.

Indeed, the values of the electron temperature typically derived from the spectral fits to the hard
spectral component in accreting black holes,  $kT_e\sim 50-150$ keV, are comfortably within the 
range defined by the two virial temperatures, but, unlike in the case of optically thick accretion solutions,
it has proved impossible to derive it from the first principles of accretion theory, and various 
models have been put forward to explain it \citep{narayan:94,blandford:99}. 

In most cases, the observed hard X-ray spectral component from hot optically thin plasma
is believe to be produced by
unsaturated Comptonization of low frequency seed photons from the accretion disc itself (when present),
with 
characteristic temperature $T_{\rm bb}$. Such a spectrum has a nearly 
power law shape in the energy range from $\sim 3 kT_{\rm bb}$ to $\sim kT_e$ \citep{st80}. 
For the parameters typical for black holes in AGN
 this corresponds to the energy range from $\sim$ a few tens of eV  to $\sim
50-100$ keV. 
The photon index $\Gamma$ of the Comptonized spectrum depends in a
rather complicated way on the parameters of the Comptonizing media, primarily on 
the electron temperature and the Thompson optical depth  \citep{st80}. In fact, 
the emerging power law slope depends more directly on the Comptonization parameter $y$, which
is set by  the energy balance in the optically thin medium: critical is
the ratio of the energy deposition rate into hot electrons and the energy flux brought into the
Comptonization region by soft seed photons \citep{st89,Dermer:91,haardt:93}.

Broadly speaking, significant part of, if not the entire diversity of the spectral behavior observed in accreting 
black holes of stellar mass can be explained by the changes in the proportions in which the gravitational  
 energy of the accreting matter is dissipated in the optically thick and optically thin parts of the accretion flow.
 This is less so for supermassive black holes in AGN, where emission sites other than the accretion disk and hot corona
 may play significant role (e.g.broad and narrow emission line regions or dusty obscuring structures on pc scales, 
see section~\ref{sec:adisc} \ref{sec:torus} above and below). 
 The particular mechanism driving these changes is however unknown. Despite significant progress in MHD
 simulations of the accretion disk achieved in recent years \citep{ohsuga:11,schnittman:13,jiang:14} there is no
 accepted global model of accretion onto a compact object able to fully explain all the different spectral 
energy distributions observed, nor the transitions among them. 

Generically, the X-ray spectra of luminous AGN are all dominated 
by a power-law in the 2-10 keV energy range, with a relative narrow distribution
of slopes: $\langle \Gamma \rangle = 1.8 \pm 0.2$ \citep{Nandra1994,steffen:06,young:10}, 
consistent with the expectations of Comptonization models discussed above, and suggesting a quite robust 
mechanism is in place to guarantee a almost universal balance between heating and cooling in the 
hot plasma. Of course, X-ray spectra of AGN are more complex than simple power-laws.
A clear reflection component from cold material 
\citep{george:91,reynolds:98} is observed in a number
of nearby AGN, and further required by Cosmic X-ray Background (CXRB) synthesis models \cite{gilli:07};
and emission and absorption features are also seen in good quality spectra. 
The most prominent and common of those
is a narrow iron K$\alpha$ emission line. Such a line, produced by cold, distant material 
appears to be dependent on luminosity, with more luminous sources having smaller equivalent widths 
\citep[the so-called Iwasawa-Taniguchi effect][]{iwasawa:93}.

The physical origin of the tight coupling between cold and hot phases, however, remains elusive. In fact, the main open questions
regarding the origin of the X-ray emitting coronae and the reflection component in AGN (and in XRB) are intimately connected with
those left open by the classical theory of relativistic accretion discs. 
The main ones concern a) the physical nature of the viscous stresses and their scaling with local 
quantities within the disc (pressure, density); b) the exact vertical structure of the 
disc and the height where most of the dissipation takes place and c) the nature of the inner boundary 
condition.  

As usual, observational hints on the right answers to those questions come more easily from
well sampled observations of transient black holes in XRB, where the dynamical evolution of the coupled
disc-corona system can be followed in great detail, proving at least a phenomenological
framework for how optically thin and optically thick plasma share accretion energy at
different accretion rates \citep{Fender:04}.

\begin{figure*}
\centering
\begin{tabular}{cc}
\includegraphics[width=0.49\textwidth]{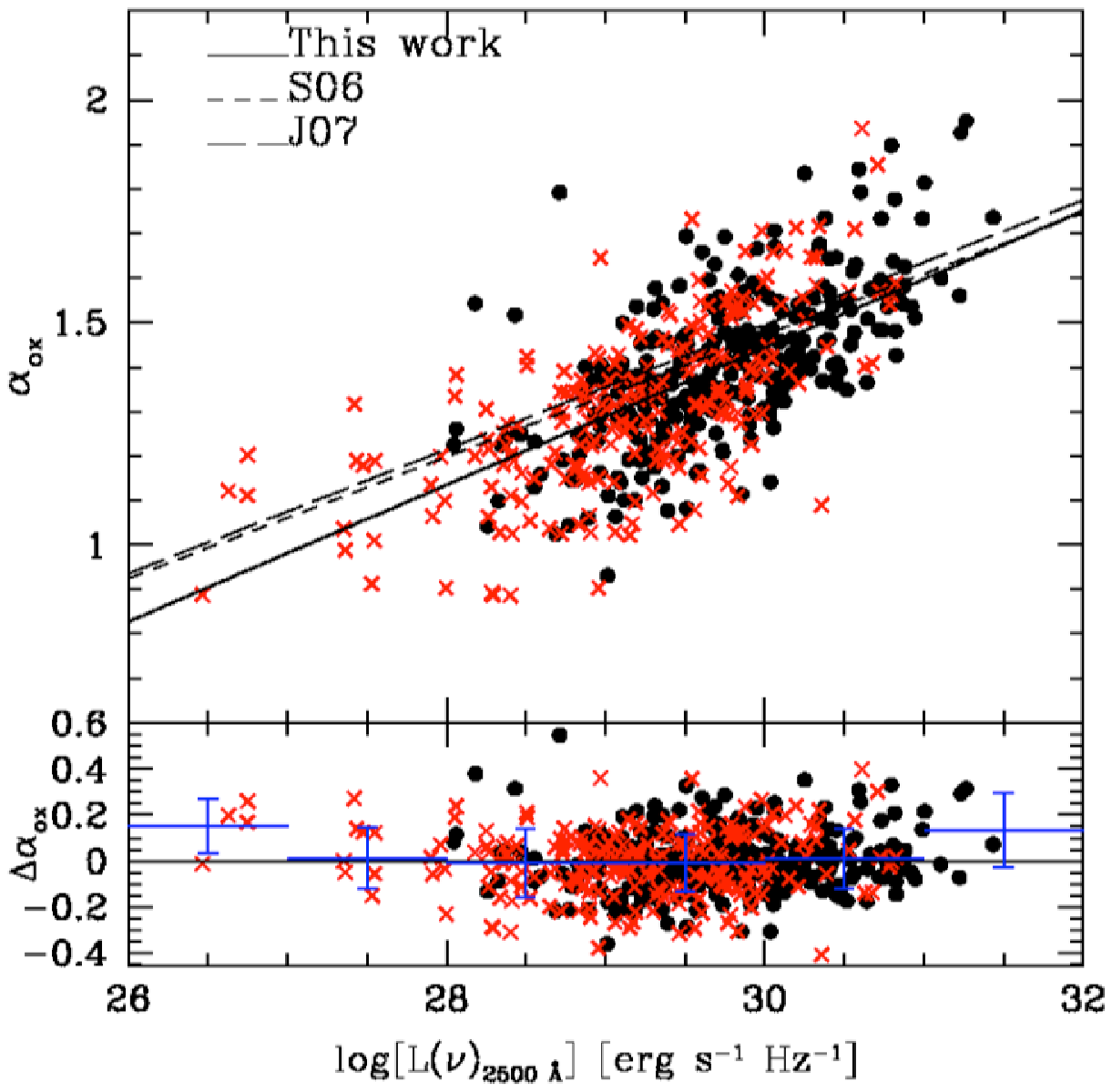}&
\includegraphics[width=0.49\textwidth]{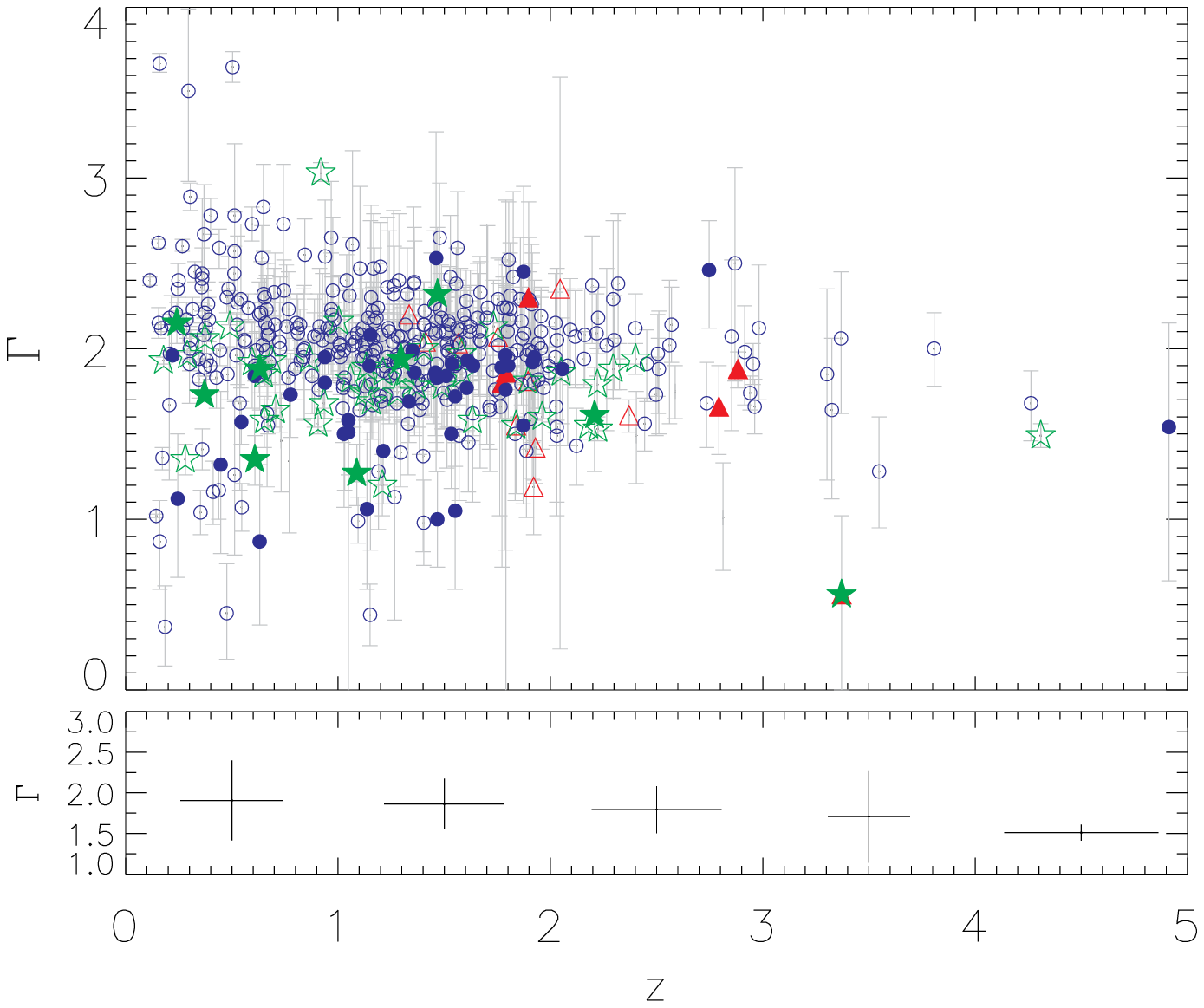}\\
\end{tabular}
\caption{{\it Left:} The optical-to-X-ray spectral slope $\alpha_{\rm ox}$ as a function of
   luminosity density at 2500$\AA$. From \citet{lusso:10}. {\it
    Right}: X-ray photon index ($\Gamma$) vs.~redshift $z$. Blue
  circles represent radio quiet, non-BAL (Broad
  Absorption Line) quasars, green stars
  represent radio loud quasars, and red triangles represent BAL
  quasars.  The bottom plot shows the weighted mean
  $\Gamma$ values for bins of width $\Delta z=1$. No clear sign of
  evolution in the average X-ray spectral slope of AGN is detected
  over more than 90\% of the age of the universe. From
  \citet{young:10}.}
\label{fig:coronae_agn}
\end{figure*}

In the case of AGN, because of the complexities of galactic nuclei discussed above, 
and because the discs and coronae
of AGN emit in distinct parts of the electromagnetic spectrum, 
it is much more difficult to clearly distinguish between
different spectral states in terms of a simple power ratios between the two main spectral components.
Nevertheless,  as a very general diagnostic, the
``X-ray loudness'', usually characterized by the $\alpha_{\rm ox}$
parameter, i.e. the slope of the spectrum between 2500 $\AA\,=5$ eV
and 2 keV: $\alpha_{\rm ox}= 0.3838 \log(F_{\rm 2keV}/F_{\rm 2500})$
can be used to characterize the fraction of bolometric light carried
away by high-energy X-ray photons.  Recent studies of large samples of
both X-ray and optical selected AGN have clearly demonstrated that
$\alpha_{\rm ox}$ is itself a function of UV luminosity, with less luminous objects being more X-ray bright 
\citep[see e.g.][and the left panel of Figure~\ref{fig:coronae_agn}]{steffen:06,lusso:10,jin:12}. 
In very general terms, this might point towards a connection between
accretion disc physics and the mechanism(s) of coronae generation in AGN
\citep{merloni:03,wang:04}.

Such generic properties of AGN X-ray SED do not appear to change significantly with redshift:
even for the most distant objects known
where reliable spectral analysis of AGN can be performed, no clear
sign of evolution in either $\alpha_{\rm ox}$ (at fixed luminosity) or 
the X-ray spectral slope $\Gamma$ has been
detected (see the right panel of Figure~\ref{fig:coronae_agn}).

\subsection{Infrared dust emission from AGN: the link with the nuclear structure at the Bondi radius}
\label{sec:torus}

The observational appearance of an AGN is not only determined by the
intrinsic emission properties of its accretion disc and corona, but also by the nature, amount, dynamical and kinematic state
of any intervening material along the line of sight. Intrinsic obscuration does indeed play
 a fundamental role for our understanding of the overall properties of AGN. As we have seen in the previous section,
the intrinsic shape of the X-ray continuum can
be characterized by a power-law in the 2-10 keV energy range, with a relative narrow distribution
of slopes: $\langle \Gamma \rangle = 1.8 \pm 0.2$.
Thus, the hard slope of the Cosmic X-ray Background (CXRB) spectrum 
(well described by a power-law with photon
index $\Gamma_{\rm CXRB} \simeq 1.4$ at $E < 10$ keV), and the prominent peak observed 
at about 30 keV are best accounted
for by assuming that the majority of active galactic nuclei are in fact obscured 
\citep[][see also Figure~\ref{fig:cxrb} below]{Setti1989,Comastri1995}.

In the traditional 'unification by orientation' schemes, 
the diversity of AGN observational classes is explained on the basis of the line-of-sight
orientation with respect to the axis of rotational symmetry of the system \citep{Antonucci1985,Antonucci1993,Urry1995}.
In particular, obscured and un-obscured AGN are postulated to be intrinsically
the same objects, seen from different angles with respect to a dusty, large-scale, 
possibly clumpy, parsec-scale medium, which 
obscures the view of the inner engine \citep{elitzur:08,netzer:08}.
According to the simplest interpretations
of such unification schemes, there should not 
be any dependence of the obscured AGN fraction with intrinsic luminosity and/or redshift.

However, the results on the statistical properties of obscured AGN from these studies are 
at odds with the simple 'unification-by-orientation' scheme. 
In fact, evidence has been mounting over the years that
the fraction of absorbed AGN, defined in different and often
independent ways, appears to be lower at higher nuclear luminosities
\citep{Lawrence1982,Ueda2003,Steffen2003,Simpson2005,Hasinger2008,Brusa2010,Bongiorno2010,Burlon2011,Assef2013,buchner:15}. 
Such an evidence, however, is not uncontroversial. As recently summarized by \citet[][and references therein]{Lawrence2010},
the luminosity dependence of the obscured AGN fraction, so clearly detected, especially in X-ray selected samples,
is less significant in other AGN samples, such as those selected on the basis of their extended, low frequency
radio luminosity \citep{Willott2000} or in mid-IR colors \citep{Lacy2007}. The reasons for these discrepancies are
still unclear, with \citet{Mayo2013} arguing for a systematic bias in the X-ray selection due to 
an incorrect treatment of complex, partially-covered AGN.

Evidence for a redshift evolution of the obscured AGN fraction is even
more controversial. Large samples of X-ray selected objects
have been used to corroborate claims of positive evolution of the fraction of
obscured AGN with increasing redshift \citep{Lafranca2005,Treister2006a,Hasinger2008}, as well as 
counter-claims of no significant evolution \citep{Ueda2003,Tozzi2006,gilli:07}. 
More focused investigation on specific AGN sub-samples,
such as $z>3$ X-ray selected QSOs \citep{Fiore2012,Vito2013}, 
rest-frame hard X-ray selected AGN \citep{Iwasawa2012}, or 
Compton Thick AGN candidates \citep{Brightman2012} 
in the CDFS have also suggested an increase of the incidence of
nuclear obscuration towards high redshift. Of critical importance 
is the ability of disentangling luminosity and redshift effects in (collections of)
flux-limited samples and the often complicated selection effects at high redshift, both in terms of 
source detection and identification/follow-up.

In a complementary approach to these ``demographic'' studies (in which the incidence of obscuration
and the covering fraction of the obscuring medium
is gauged statistically on the basis of large populations), SED-based investigations
look at the detailed spectral energy distribution of AGN, and at the IR-to-bolometric flux ratio in particular,
to infer the covering factor of
the obscuring medium in each individual source \citep{Maiolino2007,Treister2008,Sazonov2012,Roseboom2013,Lusso2013}. These studies
 also found general trends of decreasing covering factors with increasing nuclear (X-ray or bolometric) luminosity, and
little evidence of any redshift evolution \citep{Lusso2013}. Still, the results of these SED-based
investigations are not always in quantitative agreement with the demographic ones. 
This is probably due to the combined effects of the uncertain physical properties (optical depth, 
geometry and topology) of the obscuring medium \citep{Granato1994,Lusso2013}, 
as well as the unaccounted for biases in the observed {\em distribution} 
of covering factors for AGN of any given redshift and luminosity \citep{Roseboom2013}.
 To account for this, 
different physical models for the obscuring torus
have been proposed, all including some form of radiative coupling between the central
AGN and the obscuring medium \citep[e.g.][]{Lawrence1991,Maiolino2007,Nenkova2008}.

Irrespective of any specific model, it is clear that a detailed physical assessment of the interplay 
between AGN fuelling, star formation and
obscuration on the physical scales of the obscuring medium 
is crucial to our understanding of the mutual
influence of stellar and black hole mass growth in galactic nuclei \citep{Bianchi2012}.
Conceptually, we can identify three distinct spatial regions in the nucleus of a galaxy
on the basis of the physical properties of the AGN absorber. The outermost one is the gravitational
sphere of influence of the supermassive black hole (SMBH) 
itself, also called Bondi Radius 
$R_{\rm B} = 2GM_{\rm BH}/\sigma^2\simeq 10\, M_{\rm BH,8}\, \sigma_{\rm ,300}\; {\rm pc}$, where $M_{\rm BH,8}$ is the black hole mass in units of $10^{8}M_{\odot}$, and
$\sigma_{\rm ,300}$ can be either the velocity dispersion of stars for a purely collisionless nuclear environment, or
the sound speed of the gas just outside $R_{\rm B}$, measured in units of 300 km/s. 
To simplify, one can consider any absorbing
gas on scales larger than the SMBH sphere of influence to be ``galactic'', in the sense that
its properties are governed by star-formation and dynamical processes operating at the galactic scale. 
The fact that gas in the host galaxy can obscure AGN is not only predictable, but also clearly observed, 
either in individual objects (e.g. nucleus-obscuring dust lanes, \citealt{Matt2000}), or in larger samples
showing a lack of optically selected AGN in edge-on galaxies \citep{Maiolino1995,Lagos2011}.
Indeed, if evolutionary scenarios are to supersede the standard 
unification by orientation scheme and obscured AGN truly represent
a distinct phase in the evolution of a galaxy, then we expect
a relationship between the AGN obscuration distribution and the
larger scale physical properties of their host galaxies.

Within the gravitational sphere of influence of a SMBH, the most critical scale is the radius
within which dust sublimates under the effect of the AGN irradiation. A general treatment
of dust sublimation was presented in \citet{Barvainis1987,Fritz2006}, and subsequently applied to sophisticated
clumpy torus models \citep{Nenkova2008} or to interferometric observations of galactic nuclei in the near-IR
\citep{Kishimoto2007}. For typical dust composition, the dust sublimation radius is expected to scale as  
$R_{\rm d}\simeq 0.4\, (L_{\rm bol,45}/10^{45})^{1/2}(T_{\rm sub}/1500K)^{-2.6}\; {\rm pc}$ \citep{Nenkova2008}, as indeed
confirmed by interferometry observations of sizable samples of both obscured and un-obscured AGN in the nearby
Universe \citep{Tristram2009,Kishimoto2011}. 
Within this radius only atomic gas can survive, and reverberation mapping measurements do suggest that indeed
the Broad emission Line Region (BLR) is located immediately inside $R_{\rm d}$ \citep{Netzer1993,Kaspi2005}.
 
The parsec scale region between $R_{\rm d}$ and $R_{\rm B}$ is the traditional location of the obscuring torus
of the classical unified model. On the other hand, matter within $R_{\rm d}$ may be dust free, 
but could still cause substantial
obscuration of the inner tens of Schwarzschild radii of the accretion discs, where 
the bulk of the X-ray emission is produced \citep{Dai2010,Chen2012}. Indeed, a number of
X-ray observations of AGN have revealed in recent years the evidence for gas absorption within the
sublimation radius. Variable X-ray absorbers on short timescales are quite common 
\citep{Risaliti2002,Elvis2004,Risaliti2007, Maiolino2010}, and the variability timescales clearly suggests
that these absorbing structures lie within (or are part of) the BLR itself.

\begin{figure*}
\includegraphics[width=0.49\textwidth,clip]{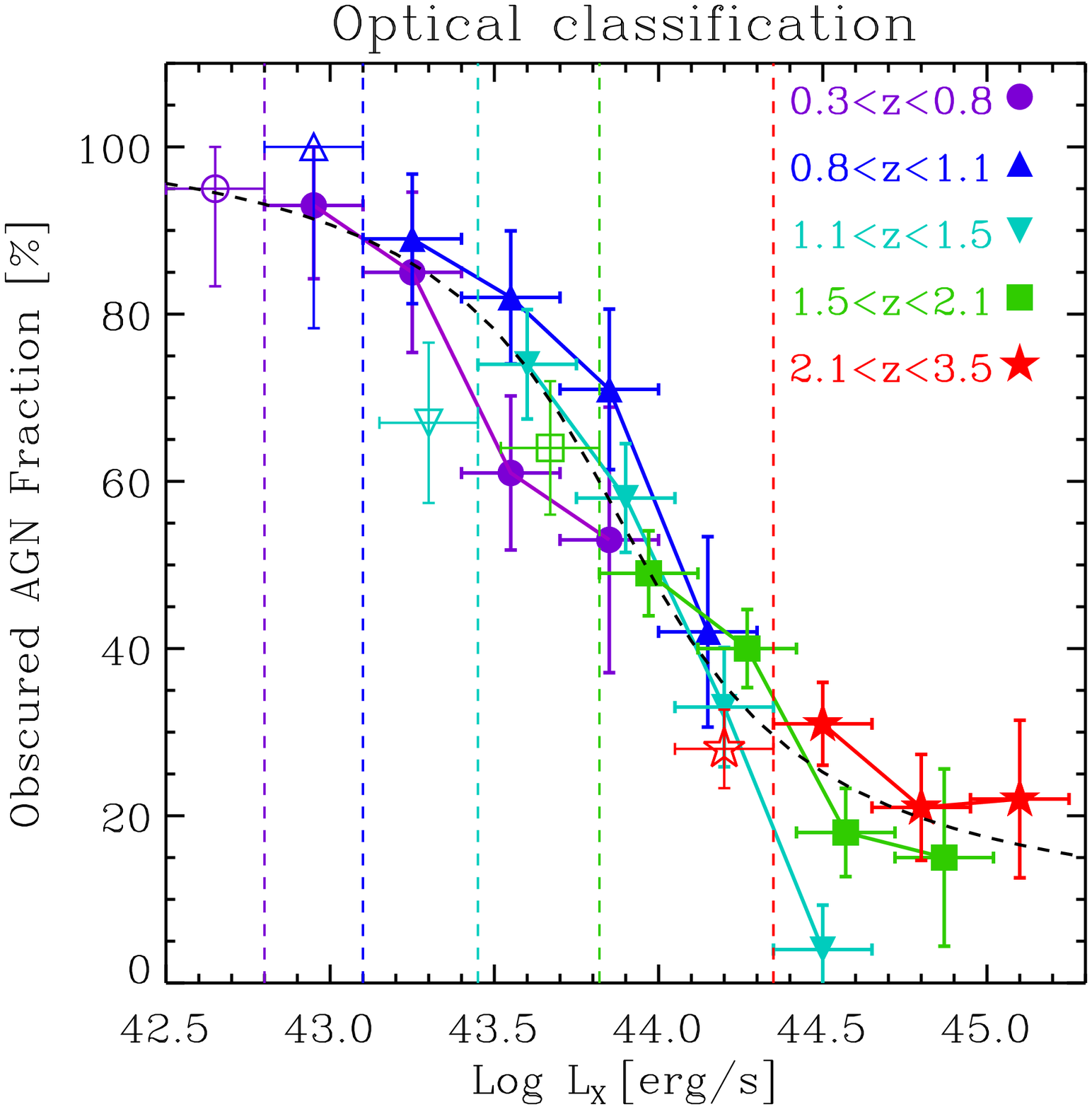}
\includegraphics[width=0.49\textwidth,clip]{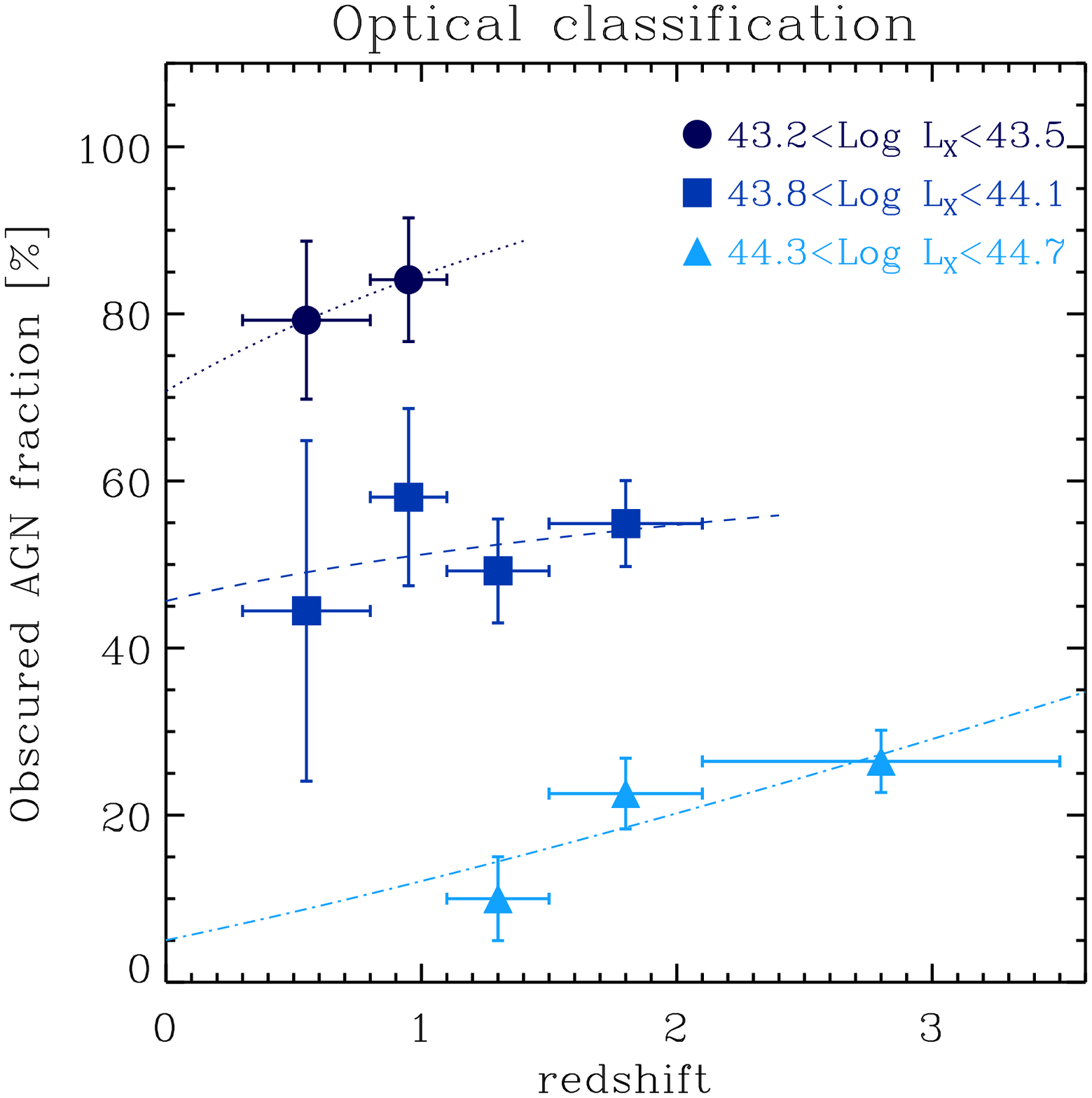}
\caption{{\it Left Panel}: The fraction of optically obscured AGN is plotted
  versus the X-ray luminosity for different redshift bins (purple
  circles: $0.3 \le z<0.8$; blue upwards triangles: $0.8 \le z<1.1$;
  cyan downwards triangles: $1.1 \le z<1.5$; green squares: $1.5 \le
  z<2.1$ and red stars: $2.1 \le z<3.5$). The vertical dashed lines
  mark the luminosities above which the samples are complete in each
  redshift bin (of corresponding color). Empty symbols are from
  incomplete bins. The
  dashed line is the best fit
  to the entire data set across the whole redshift range.  
{\it Right Panel}: Redshift evolution of the fraction of Obscured AGN in different luminosity bins (only those for which we are 
complete have been shown). The dotted, dashed, and dot-dashed lines show the best fit evolution in the three luminosity interval, respectively.} 
\label{fig:abs_frac_lx_allz}
\end{figure*}

In \cite{merloni:14} we have examined the luminosity and redshift dependence of the fraction of AGN
classified as obscured, both optically and from the X-ray spectra. The sample, X-ray selected in the XMM-COSMOS
field, was selected on the basis of the estimated {\em rest frame} 2-10 keV flux, in order to avoid as much as possible biases
due to the $z-N_{\rm H}$ degeneracy for obscured AGN.

The left hand panel of 
Figure~\ref{fig:abs_frac_lx_allz} shows such a fraction
as a function of intrinsic X-ray luminosity in five different redshift bins. 
The decrease of 
the obscured AGN fraction with luminosity is strong, and confirms previous studies on the XMM-COSMOS 
AGN \citep{Brusa2010}. 
The dashed line, instead, shows the best fit relations to the
optically obscured AGN fraction obtained combining all redshift bins. 

\begin{figure*}
\includegraphics[width=0.49\textwidth,clip]{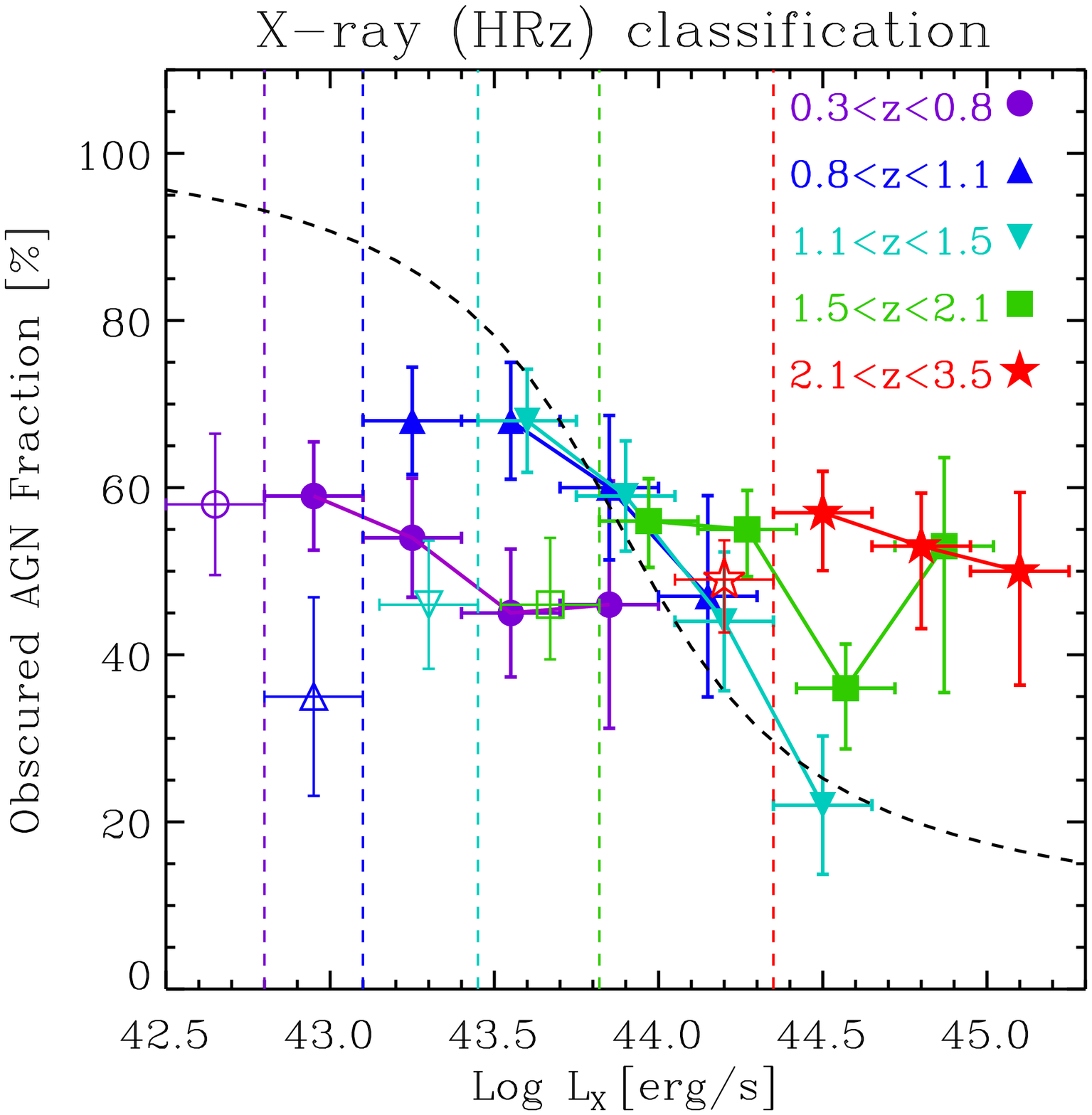}
\includegraphics[width=0.49\textwidth,clip]{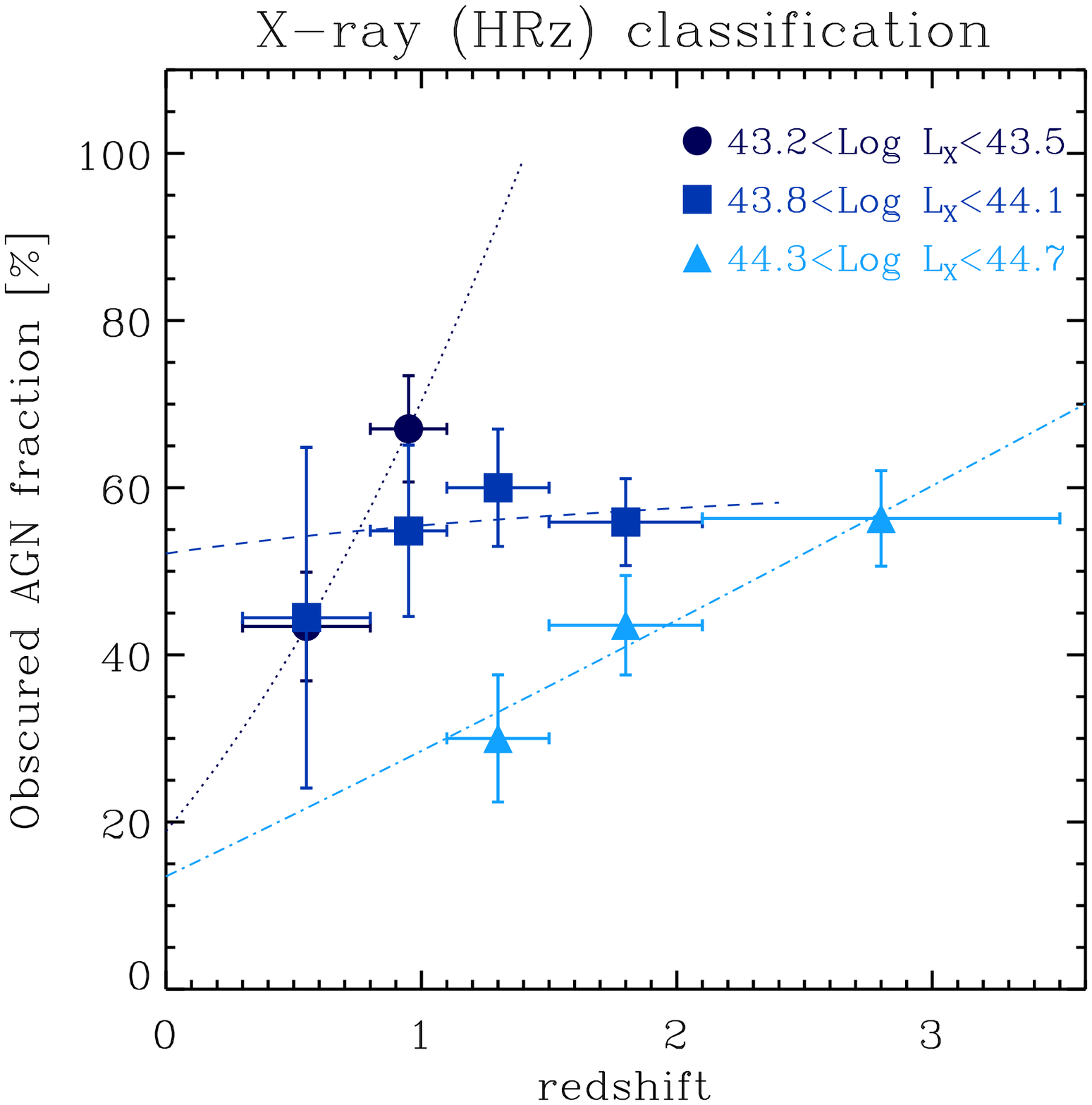}
\caption{{\it Left panle:} The fraction of X-ray obscured AGN is plotted
  versus the X-ray luminosity for different redshift bins (purple
  circles: $0.3 \le z<0.8$; blue upwards triangles: $0.8 \le z<1.1$;
  cyan downwards triangles: $1.1 \le z<1.5$; green squares: $1.5 \le
  z<2.1$ and red stars: $2.1 \le z<3.5$). The vertical dashed lines
  mark the luminosities above which the samples are complete in each
  redshift bin (of corresponding color). Empty symbols are from
  incomplete bins. The dashed line is here plotted as a reference, and represent the best fit to the absorbed AGN fraction vs. luminosity
relation for optically obscured AGN, from Fig.~\ref{fig:abs_frac_lx_allz}.
{\it Right Panel}: Redshift evolution of the fraction of Obscured AGN in different luminosity bins (only those for which we are 
complete have been shown). The dotted, dashed, and dot-dashed lines show the best fit evolution in the three luminosity interval, 
respectively.} 
\label{fig:abs_frac_lx_allz_hrz}
\end{figure*}

One of the most important conclusions of the work of Merloni et al. (2014) was that for about 30\% of all X-ray selected AGN
the optical- and X-ray-based classifications into obscured and un-obscured sources disagree. 
For this reason, the left panel of Figure~\ref{fig:abs_frac_lx_allz_hrz}, which shows the fraction of (X-ray classified) 
obscured AGN as a function
of intrinsic 2-10 keV X-ray luminosity ($L_{\rm X}$) is significantly different from that of Figure~\ref{fig:abs_frac_lx_allz}. 
In particular, at low luminosity
about one third of the AGN have un-obscured X-ray spectra but no broad emission lines or prominent blue accretion disc continuum in their optical spectra, while, on the other hand, about 30\% of the most luminous QSOs have obscured X-ray spectra despite showing clear 
broad emission line in the optical spectra.

We plot in the right panels of 
Figures~\ref{fig:abs_frac_lx_allz} and \ref{fig:abs_frac_lx_allz_hrz} the fraction of obscured AGN as a function of
redshift, for three separate luminosity intervals and for the optical and X-ray classifications, respectively. 
Only bins where the sample is complete are shown.
For optically classified AGN, we do not see any clear redshift evolution, apart from the highest luminosity objects 
(i.e. genuine QSOs in the XMM-COSMOS sample, with $L_{\rm X}$ between 10$^{44.3}$ and 10$^{44.7}$ erg/s). To better quantify this,
we have fitted separately the evolution of the obscured fraction in the three luminosity bins with the function:
$F_{\rm obs}=B \times (1+z)^{\delta}$.

The best fit relations are shown as thin lines in
the right panels of Figures~\ref{fig:abs_frac_lx_allz} and 
\ref{fig:abs_frac_lx_allz_hrz}. For the optical classification, as anticipated, 
we measure a significant evolution ($\delta_{\rm OPT} > 0$) 
only for the most luminous objects, with $\delta_{\rm OPT}=1.27\pm0.62$. 
For the X-ray classification, we observe a significant evolution with redshift 
both  at the lowest and highest luminosities, where the fraction of X-ray obscured AGN increases 
with $z$, consistent with previous findings
by numerous authors
 \citep[][]{Lafranca2005,Treister2006a,Hasinger2008,Fiore2012,Vito2013}. 
A more robust assessment of the redshift evolution 
of the obscured AGN fraction, however, would require a more extensive coverage of the $L-z$ plane than that afforded by the 
flux-limited XMM-COSMOS sample. 

\subsection{The SED of low luminosity AGN}
\label{sec:llagn}

\begin{figure}
\centering
\hbox{
\includegraphics[width=\textwidth]{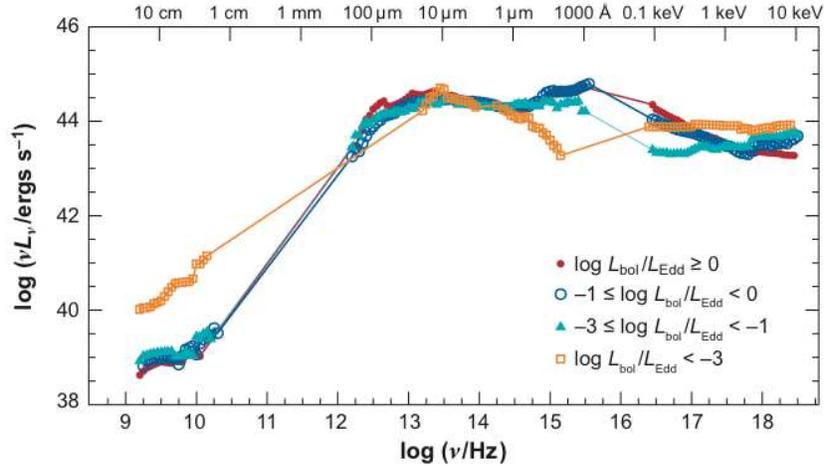}
\hspace{0.2cm}
}
\caption{Energy spectra of a
 compilation of AGN at different Eddington-scaled luminosities (from \cite{ho:08}).}
\label{fig:states_spe}       
\end{figure}

Irrespective of where and when was most of the mass in SMBH accreted 
(and we will see in section~\ref{sec:soltan} that 
this happened most likely at
high accretion rates), the steepness of the AGN luminosity function tells us that most of the {\em time} in the life
of a nuclear black hole is spent in a
low accretion rate, low luminosity regime: 
the ubiquity of SMBH in the nuclei of nearby galaxies implies that, in the local Universe, 
AGN of low and very low luminosity  vastly outnumber their bright and active counterparts. 
However, at lower accretion rates the precise determination of AGN SED is severely hampered by the contamination from 
stellar light (as we discussed in more detail in section~\ref{sec:surveys_agn}), 
and very high resolution imaging is needed to identify the accretion-related emission.
This is of course possible only for nearby galaxies, and only a limited number of reliable SED of LLAGN are currently 
known \cite{ho:08}.
An important step towards the classification of AGN in terms of their specific modes of 
accretion was taken by \cite{mhd03} and \cite{falcke:04}, whereby a ``fundamental plane'' relation between mass, X-ray
and radio core luminosity of active black holes was discovered and characterized in terms of accretion flows.
In particular, the observed scaling between radio luminosity, X-ray luminosity and BH mass implies that the 
output of low-luminosity AGN is dominated by
kinetic energy rather than by radiation \citep{sams96,merloni:07,hardcastle:07,best:12}. 
This is in agreement with the average SED of LLAGN
\citep{ho:08,nemmen:14} displaying a clear lack of thermal (BBB) emission associated to 
an optically thick accretion disc, strongly suggestive of a ``truncated disc'' scenario and of a radiative inefficient
inner accretion flow (see Figure~\ref{fig:states_spe}). 

\section{AGN luminosity functions and their evolution}
\label{sec:lf}

The Luminosity Functions (LF) $\phi(L)$ describe the space density of sources of different luminosity $L$, so that
$dN= \phi(L) dL$ is the number of sources per unit volume with luminosity in the range $L + dL$. In this section
we review the current observational state of affairs in the study of AGN luminosity functions, in various parts of the 
electromagnetic spectrum (from radio to X-rays). 
Well constrained single-band luminosity functions, together with a good understanding
of the AGN SED (and its evolution), can then be used to infer the ``bolometric'' LF, i.e. the full inventory of the
radiative energy release onto accreting black holes.

\subsection{The evolution of radio AGN}
\label{sec:radio_lf}

The observed number counts distribution of radio sources 
\citep[see e.g.][]{merloni:13} has been used for many years as a prime tool to infer properties 
of the cosmological evolution of radio AGN, mainly because of the sensitivity of radio telescope
to distant quasar, and of the difficulty in getting reliable counterpart identification and
redshift estimate for large number of radio sources.

At bright fluxes, counts rise more steeply than the Euclidean slope $S^{-3/2}$.  This was
already discovered by the first radio surveys at meter wavelengths
\citep{ryle:55}, lending strong support
for evolutionary cosmological models, as opposed to theories of a
steady state universe.

At fluxes fainter than about a Jansky\footnote{A Jansky (named after
  Karl Jansky, who first discovered the existence of radio waves from
  space) is a flux measure, corresponding to $10^{-23}$ ergs cm$^{-2}$
  Hz$^{-1}$} (or $\approx 10^{-14}$ ergs s$^{-1}$ cm$^{-2}$ at 1 GHz)
the counts increase less steeply than $S^{-3/2}$, being
dominated by sources at high redshift, thus probing a substantial
volume of the observable universe.

At flux densities above a mJy the population of radio sources is
largely composed by AGN. For these sources, the observed
radio emission includes the classical extended jet and double lobe
radio sources as well as compact radio components more directly
associated with the energy generation and collimation near the central
engine.

\begin{figure*}
\begin{center}
\resizebox{!}{0.7\textwidth}{\includegraphics{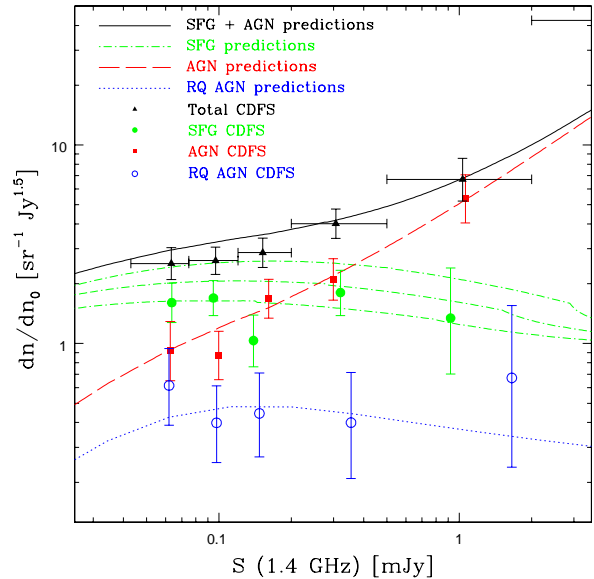}}
\end{center}
\caption{Euclidean normalized 1.4 GHz CDFS radio source counts: total counts
(black triangles), Star Forming Galaxies (filled green circles), all AGNs (red squares), and radio-
quiet AGNs (open blue circles). Error bars correspond to 1$\sigma$
errors. Model calculations refer to SFG (green dotted–dashed lines), displayed
with a 1$\sigma$
range on the evolutionary parameters, all AGNs (red dashed line),
radio-quiet AGNs (blue dotted line), and the sum of the first two (black solid
line). From \citet{padovani:09}.}
\label{fig:padovani}
\end{figure*}

The deepest radio surveys, however, (see e.g.~\citealt{padovani:09}
and references therein), probing well into the sub-mJy regime, clearly
show a further steepening of the counts. The nature of this change is
not completely understood yet, but in general it is attributed to the
emergence of a new class of radio sources, most likely that of
star-forming galaxies and/or radio quiet AGN (see Fig.~\ref{fig:padovani}).  
Unambiguous solutions
of the population constituents at those faint flux levels requires not
only identification of the (optical/IR) counterparts of such faint
radio sources, but also a robust understanding of the physical
mechanisms responsible for the observed emission both at radio and
optical/IR wavelengths.

Thus, the complex shape of the observed number counts provides clues
about the evolution of radio AGN, as well as on their physical nature,
even before undertaking the daunting task of identifying substantial
fractions of the observed sources, determining their distances, and
translating the observed density of sources in the redshift-luminosity
plane into a (evolving) luminosity function.  Pioneering work from
\citet{longair:66} already demonstrated that, in order to reproduce
the narrowness of the observed 'bump' in the normalized counts around
$1$ Jy, 
only the most luminous sources could evolve strongly with
redshift.  This was probably the first direct hint of the intimate
nature of the {\em differential} evolution AGN undergo over cosmological
times.

\begin{figure*}
\begin{center}
\resizebox{!}{0.48\textwidth}{\includegraphics{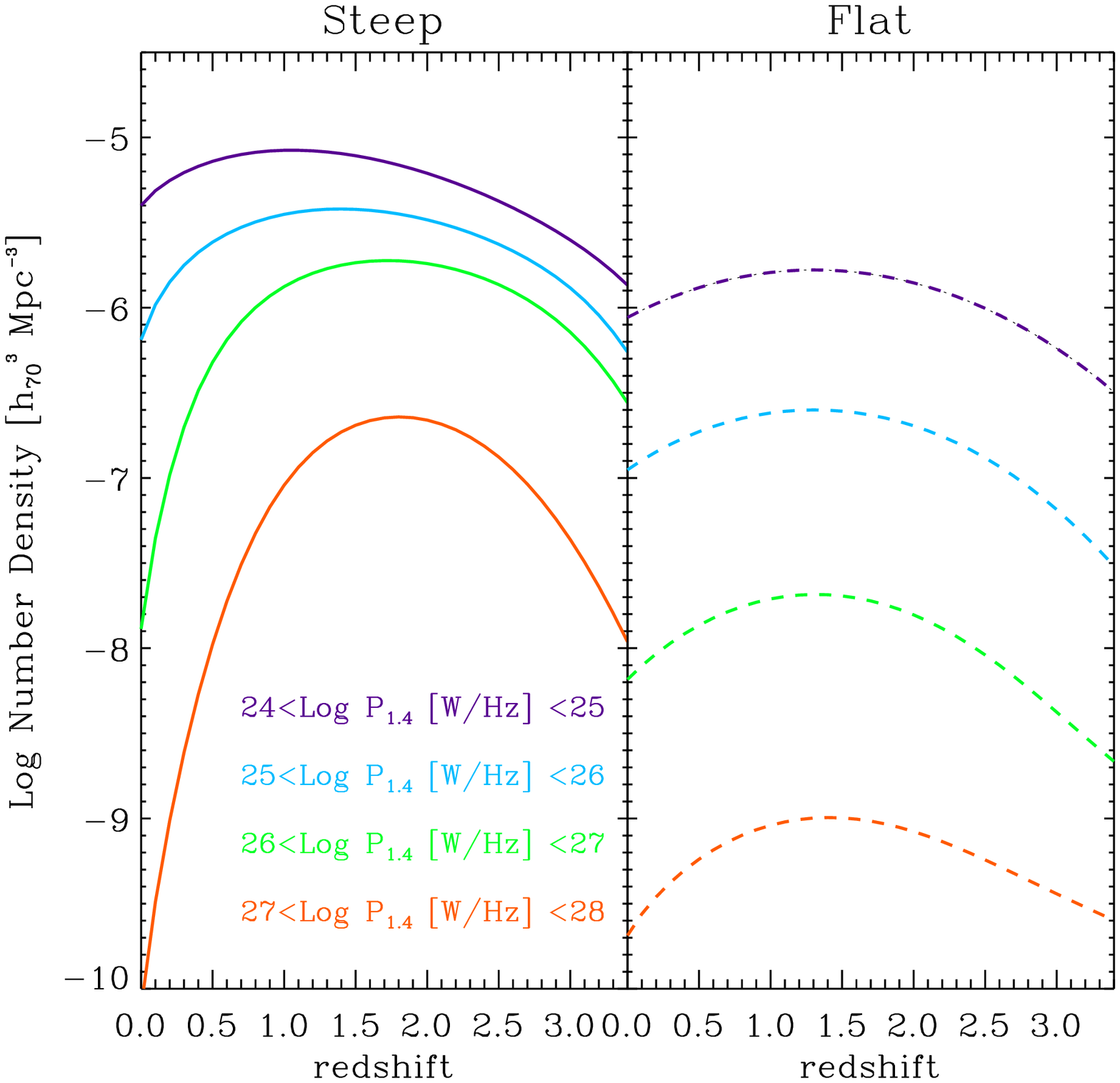}}
\resizebox{!}{0.48\textwidth}{\includegraphics{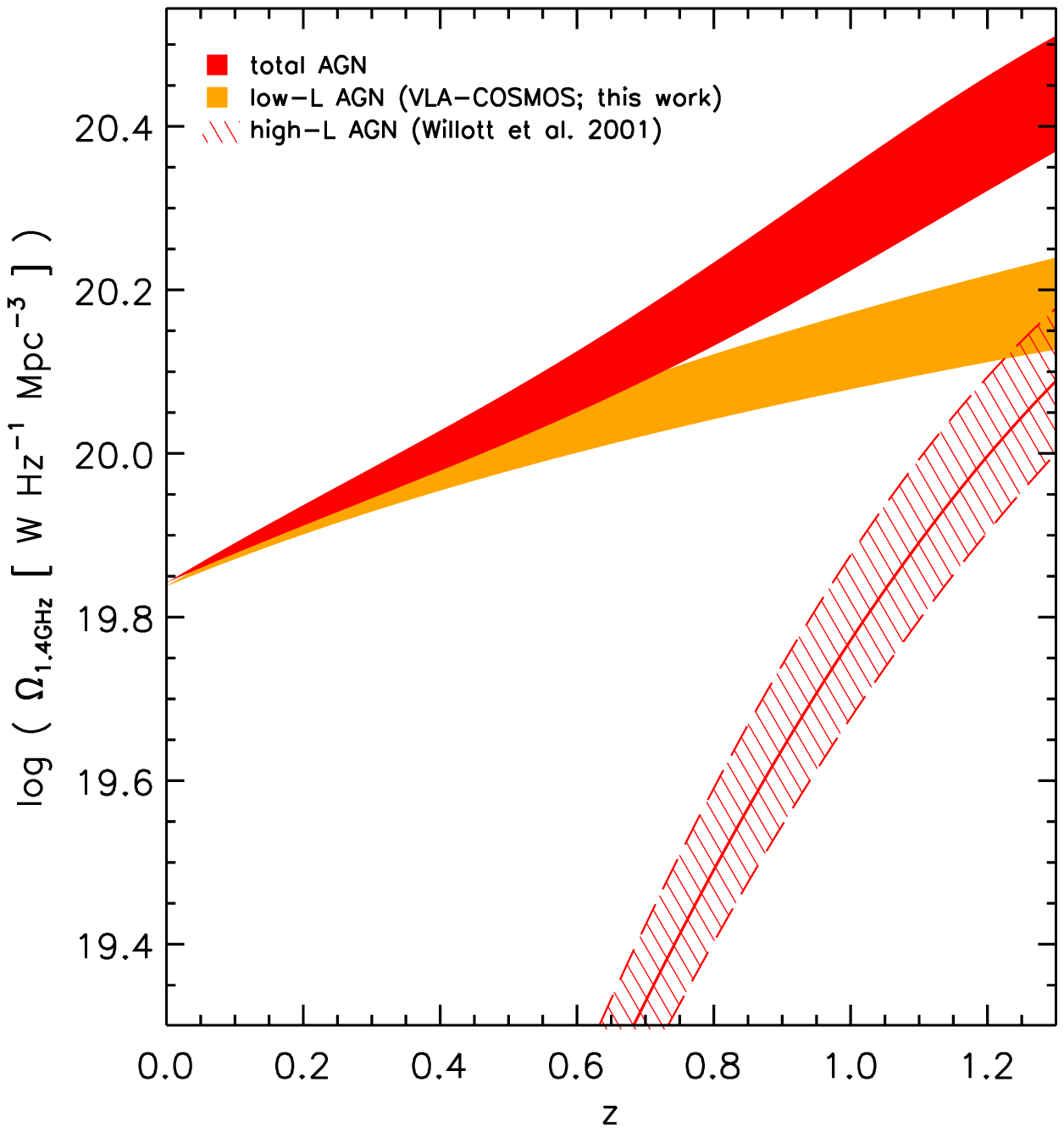}}
\end{center}
\caption{The radio view of AGN downsizing. {\it Left:} Best fit number
  density evolution of radio sources of different power, taken from
  the models of \citet{massardi:10}, for steep and flat spectrum
  sources in the left and right panels, respectively. {\it Right:}
  Evolution of the comoving 20 cm integrated luminosity density for
  VLA-COSMOS AGN (orange curve) galaxies for $z < 1.3$.  Also shown is
  the evolution of the high-luminosity radio AGN, adopted from
  (\citealt{willott:01}, hatched region; the thick and dashed lines
  correspond to the mean, maximum and minimum results, respectively).
  The evolution for the total AGN population, obtained by co-adding
  the VLA-COSMOS and high luminosity AGN energy densities, is shown as
  the red-shaded curve (adopted from \citealt{smolcic:09}). }
\label{fig:smolcic}
\end{figure*}

Indeed, many early investigations of high redshift radio luminosity functions
(see, e.g., \citealt{danese:87}) demonstrated that no simple LF evolution
models could explain the observed evolution of radio sources, with
more powerful sources (often of FRII morphology) displaying a far more dramatic rise in
their number densities with increasing redshift (see also
\citealt{willott:01}).

Trying to assess the nature of radio AGN evolution across larger
redshift ranges requires a careful evaluation of radio spectral
properties of AGN.  Steeper synchrotron spectra are produced in the
extended lobes of radio jets, while flat spectra are usually
associated with compact cores.  For objects at distances such that no
radio morphological information is available, the combination of
observing frequency, K-corrections, intrinsic source variability and
orientation of the jet with respect to the line of sight may all
contribute to severe biases in the determination of the co-moving
number densities of sources, especially at high redshift
\citep{wall:05}.

In a very extensive and equally influential work \citet{dunlop:90}
studied the evolution of the luminosity functions of steep and flat
spectrum sources separately.  They showed that the overall redshift
evolution of the two classes of sources were similar, with steep
spectrum sources outnumbering flat ones by almost a factor of ten.
Uncertainties remained regarding the possibility of a high-redshift
decline of radio AGN number densities. The issue is still under
discussion, with the most clear evidence for such a decline observed
for flat-spectrum radio QSO at $z>3$ \citep{wall:05}, consistent with
the most recent findings of optical and X-ray surveys.

Under the simplifying assumption that the overall radio AGN population
can be sub-divided into steep and flat spectrum sources, characterized
by a power-law synchrotron spectrum $S_{\nu} \propto \nu^{-\alpha}$,
with slope $\alpha_{\rm flat}=0.1$ and $\alpha_{\rm steep}=0.8$,
respectively, a redshift dependent luminosity function can be derived
for the two populations separately, by fitting simple models to a very
large and comprehensive set of data on multi-frequency source counts
and redshift distributions obtained by radio surveys at $\nu<5$ GHz
\citep{massardi:10}.  The comoving number densities in bins of
increasing radio power (at 1.4 GHz) from the resulting best fit
luminosity function models are shown in the left panel of
Figure~\ref{fig:smolcic}.

Radio AGN, both with steep and flat spectrum, show the distinctive
feature of a differential density evolution, with the most powerful
objects evolving more strongly towards higher redshift, a
phenomenological trend that, in the current cosmologist jargon, is
called ``downsizing''.

Recent radio observational campaigns of large multi-wavelength sky
surveys have also corroborated this view, by providing a much more
detailed picture of low luminosity radio AGN. For example, the work of
\citet{smolcic:09} on the COSMOS field showed that radio galaxies with
$L_{\rm 1.4 GHz} < few \times 10^{25}\,{\rm W Hz^{-1}}$ evolve up to
$z \simeq 1$, but much more mildly than their more luminous
counterparts, as shown in the right panel Figure~\ref{fig:smolcic}.

In the local Universe, the combination of the SDSS optical spectroscopic survey with the
wide-area, moderately deep {\em VLA} surveys (NVSS, \citealt{condon:98} and FIRST, \citealt{becker:95}), have been
used by \citet{best:12} to gain a powerful insight on the radio AGN population.
They found not only that the radio AGN population can be clearly distinct into two sub-groups on the basis of
their optical emission line properties (high- and low-excitations radio galaxies, HERG and LERG, respectively), 
but that dichotomy corresponds to a more profound difference in the accretion mode onto their respective black holes:
HERG typically have accretion rates between one per cent and
10 per cent of their Eddington rate, and appear to be in a radiative efficient mode of accretion, 
while most LERG accrete with an Eddington ratio of less than one per cent. In addition, the two populations show
differential cosmic evolution at fixed radio luminosity: HERG evolve strongly at all radio
luminosities, while LERG show weak or no evolution, consistent with the general trends observed in the more
distant Universe, and described above.

\subsection{Optical and Infrared studies of QSOs}

Finding efficient ways to select QSO in large optical surveys, trying
to minimize contamination from stars, white dwarfs and brown dwarfs
has been a primary goal of optical astronomers since the realization
that QSO were extragalactic objects often lying at cosmological
distances \citep{schmidt:83,richards:06}.

Optical surveys remain an extremely powerful tool to uncover the
evolution of un-obscured QSOs up to the highest redshift ($z\sim 6$).
In terms of sheer numbers, the known population of SMBH is dominated
by such optically selected AGN (e.g. more that $3\times 10^5$ QSOs have been
identified in the first three generations of the Sloan Digital Sky Survey, \citealt{paris:14}), essentially due to the
yet unsurpassed capability of ground-based optical telescopes to perform
wide-field, deep surveys of the extra-galactic sky.

\begin{figure*}
\centering
\begin{tabular}{cc}
\includegraphics[width=0.49\textwidth]{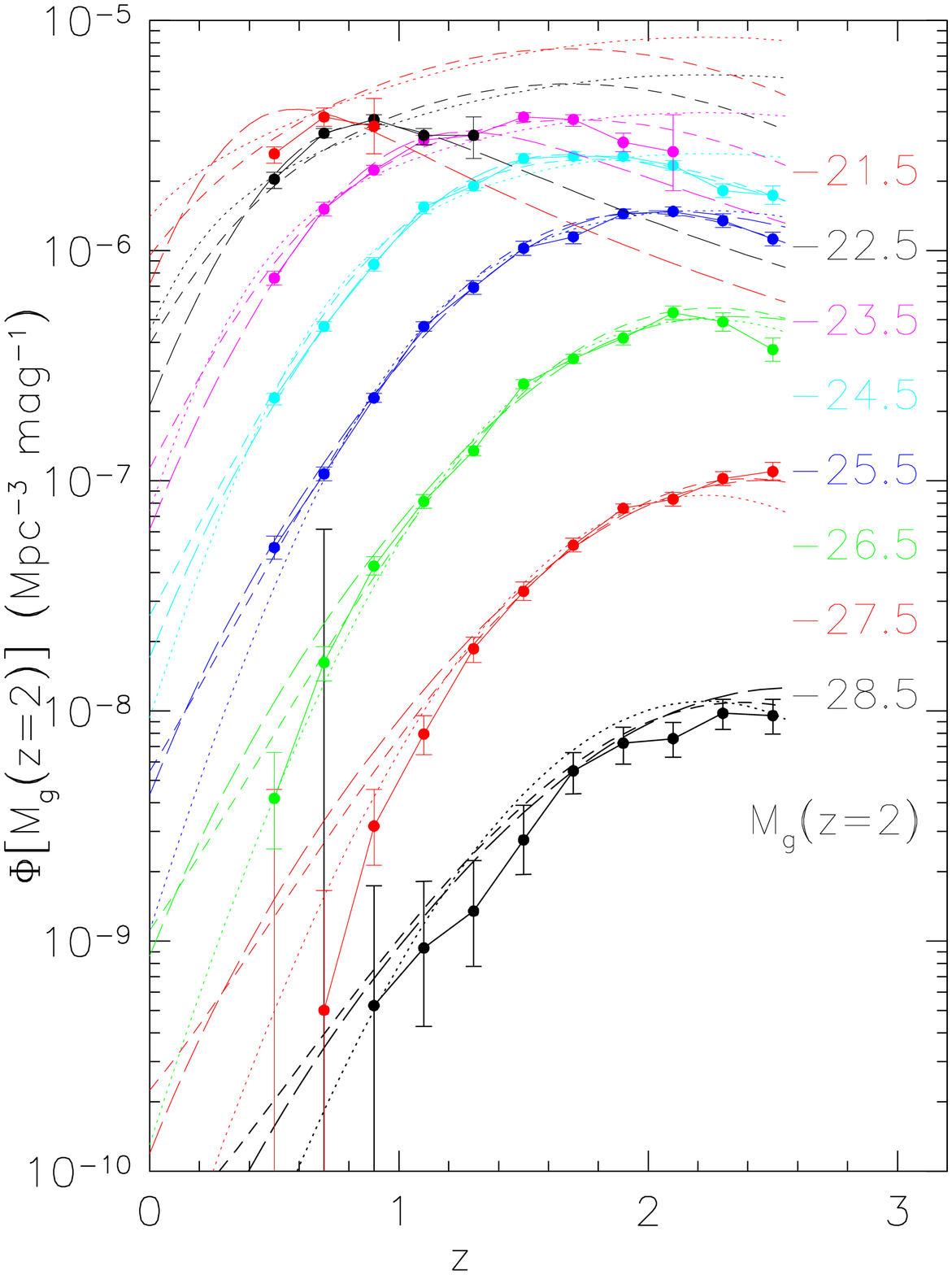}&
\includegraphics[width=0.49\textwidth,height=8cm]{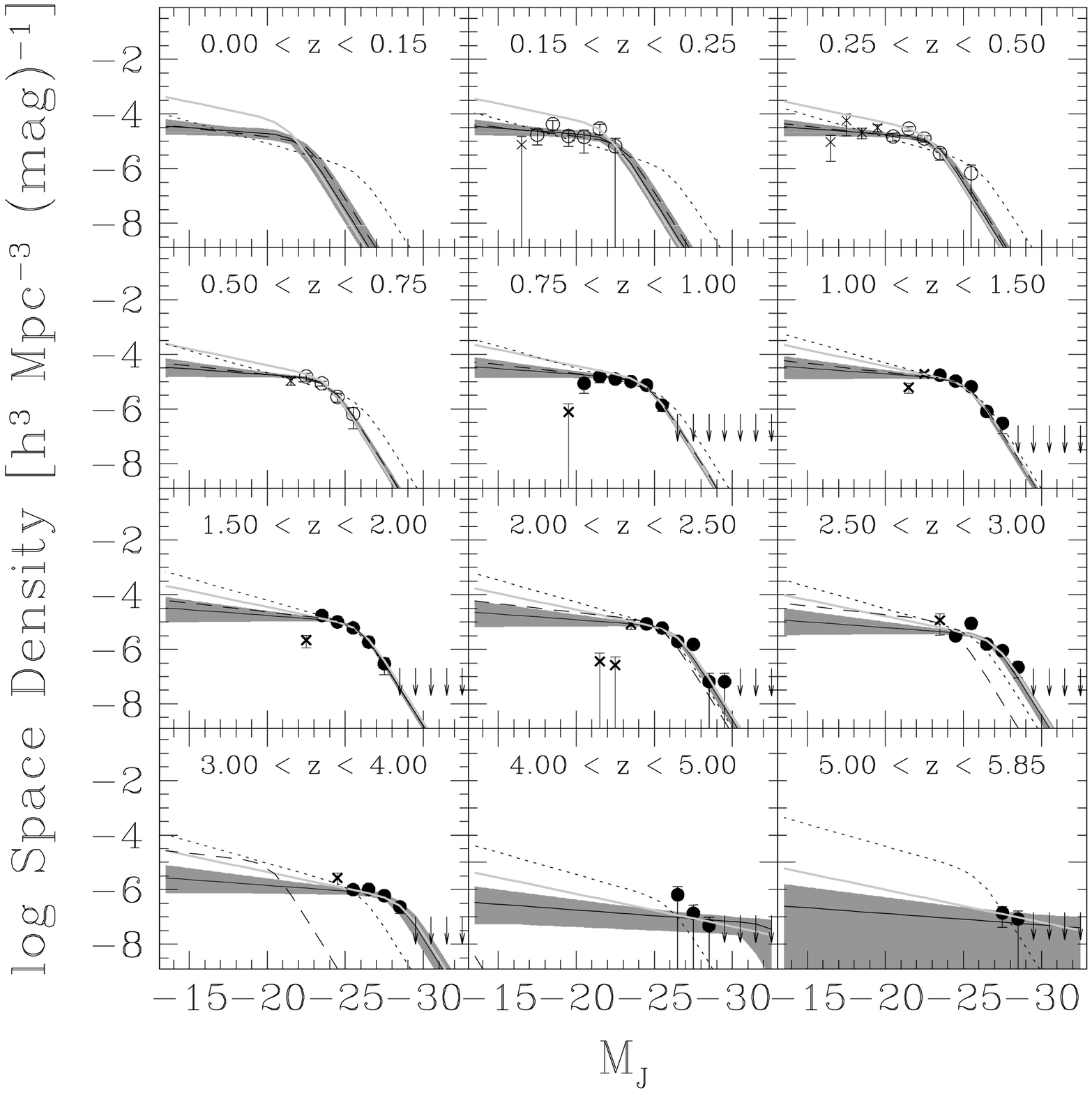}\\
\end{tabular}
\caption{{\it Left}: The combined 2SLAQ and SDSS optical QSO
  luminosity function plotted as a function of redshift for different
  absolute $g$ band magnitude intervals (the brightest at the bottom
  of the plot and the faintest at the top). The measured LF is
  compared to the best fit PLE model (Pure Luminosity Evolution, dotted lines), smooth LDDE model
  (Luminosity Dependent Density Evolution, long dashed lines) and LADE (Luminosity And Density Evolution) 
  model (short dashed lines). From \cite{croom:09}; {\it Right:} J-band luminosity function of
  mid-IR-selected AGN for several redshift bins.  The crosses show
  points that were not used in the fits.  The best-fit LADE, PLE, and
  pure PDE models are shown by the solid, dashed, and
  dotted line, respectively, although only the LADE model is an
  acceptable fit to the data.  The shaded area shows the 2$\sigma$
  confidence region for the LADE fit.  For reference, the solid light
  gray line shows the best-fit LADE model to a sample from a combined
  IR/X-ray selection.  From
  \cite{assef:11}. \label{fig:downsizing_opt_ir} }
\end{figure*}

As for the general evolution of the optically selected QSO luminosity
function, it has been known for a long time that luminous QSOs were
much more common at high redshift ($z \sim 2$). Nevertheless, it is
only with the aid of the aforementioned large and deep surveys
covering a wide enough area of the distance-luminosity plane that it
was possible to put sensible constraints on the character of the
observed evolution.  Most recent attempts \cite{croom:09} have
shown unambiguously that optically selected AGN do not evolve
according to a simple pure luminosity evolution, but instead more luminous objects peaked in
their number densities at redshifts higher than lower luminosity
objects, as shown in the left panel of
Figure~\ref{fig:downsizing_opt_ir}.

As I discussed in section~\ref{sec:torus}, according to the AGN unification paradigm obscuration comes from
optically thick dust blocking the central engine along some lines of
sight.  The temperature in this structure, which can range up to 1000K
(the typical dust sublimation temperature), and the roughly isotropic
emission toward longer wavelengths should make both obscured and
un-obscured AGNs very bright in the mid- to far-infrared bands.  This
spectral shift of absorbed light to the IR has allowed sensitive
mid-infrared observatories ({\em IRAS, ISO, Spitzer}) to deliver large
numbers of AGN \citep[see, e.g.][]{treister:06}.

Deep surveys with extensive multi-wavelength coverage have been
used to track the evolution of active galaxies in the mid-infrared
\cite[see, e.g.][]{assef:11,delvecchio:14}.  Strengthening similar conclusions
discussed above from other wavelengths, IR-selected AGN do not appear
to evolve following either a 'pure luminosity evolution'  or 'pure density evolution' 
parametrizations, but
require significant differences in the evolution of bright and faint
sources, with the number density of the former declining more steeply
with decreasing redshift than that of the latter (see the right panel
in Figure~\ref{fig:downsizing_opt_ir}).

The problem with IR studies of AGN evolution, however, lies neither in
the {\em efficiency} with which growing supermassive black holes can
be found, nor with the {\em completeness} of the AGN selection, which
is clearly high and (almost) independent of nuclear obscuration, but
rather in the very high level of {\em contamination}, as we discussed in section~\ref{sec:surveys_agn}.  
IR counts are,
in fact, dominated by star forming galaxies at all fluxes: unlike the case of the CXRB, 
AGN contribute only a small
fraction (up to 2-10\%) of the cosmic IR background radiation
\citep{treister:06}, and similar fractions are estimated for the
contribution of AGN at the ``knee'' of the total IR luminosity
function at all redshifts. This fact, and
the lack of clear spectral signatures in the nuclear, AGN-powered
emission in this band, implies that secure identification of AGN in
any IR-selected catalog often necessitates additional information from
other wavelengths, usually radio, X-rays, or optical spectroscopy.

\begin{figure}
\begin{center}
\includegraphics[width=0.7\textwidth]{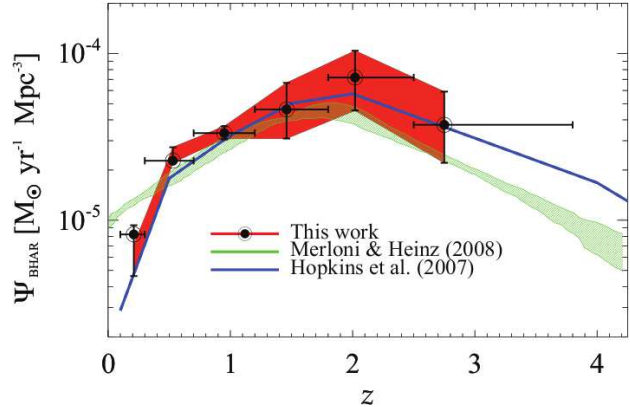}
\end{center}
\label{fig:delvecchio}
\caption{Black Hole Accretion Rate Density estimate from the {\em Herschel} selected AGN luminosity function, 
as a function of redshift (black circles). 
The red shaded area shows the $1\sigma$ uncertainty region. 
Previous estimates from different selection wavelengths (from Merloni \& Heinz 2008, and Hopkins et al. 2007) 
are reported for comparison. From Delvecchio et al. (2014).}
\end{figure}

In the best cases (such as the COSMOS and CDFS fields), 
accurate SED modelling of IR selected galaxies can be used to identify reliably AGN 
(at least those accreting at a substantial rate). \citet{delvecchio:14} have indeed been able
to use a {\em Herschel} selected sample to derive the
AGN luminosity function across a wide redshift range ($0<z<3$). Figure~\ref{fig:delvecchio} shows the total
integrated black hole accretion rate density derived from this work.

\subsection{X-ray surveys}
\label{sec:xray_surveys}
Due to the relative weakness of X-ray emission from stars and stellar
remnants (magnetically active stars, cataclysmic variables and, more
importantly, X-ray binaries are the main stellar X-ray sources), the
X-ray sky is almost completely dominated by the evolving SMBH
population, at least down to the faintest fluxes probed by current
X-ray focusing telescopes (see eq.~(\ref{eq:xlim})).

\begin{figure*}
\centering
\includegraphics[width=.8\textwidth]{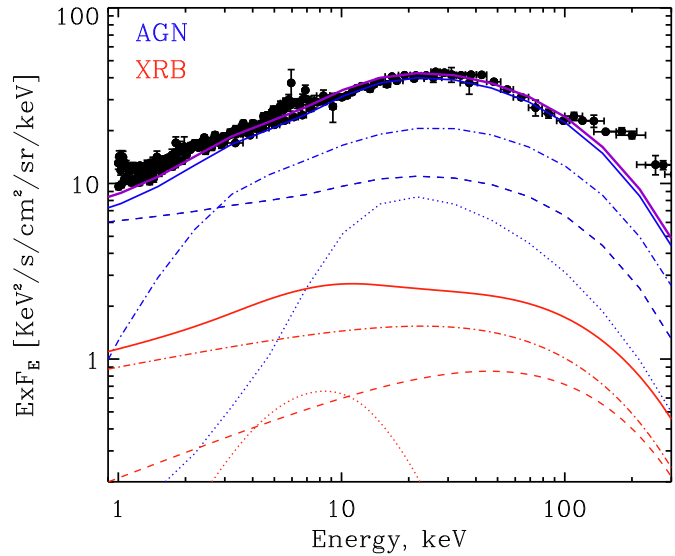}
\caption{Contribution of supermassive black holes in AGN and stellar mass black holes in X-ray binaries to the spectral 
intensity of the Cosmic X-ray background. Points with error bars show the observed CXRB brightness (as compiled from \cite{gilli:07}). 
The blue solid line is the overall contribution of AGN, which is the sum of the contributions from: i) un-obscured AGN (i.e. those with absorbing column density $N_{\rm H}<10^{22}$ cm$^2$, dashed blue line); ii) obscured, Compton-Thin AGN (with $10^{22}<N_{\rm H}<10^{24}$ cm$^2$, dot-dashed blue line) and iii) heavily obscured, Compton-Thick sources ($N_{\rm H}>10^{24}$ cm$^2$, dotted blue line). The computation is based on the \cite{gilli:07} synthesis model. 
The red solid line is the integrated contribution of high-mass X-ray binaries in star-forming galaxies, computed as described in \citet{dijkstra:12}. It is further subdivided into: i) Ultra-Luminous X-ray sources (Ultra-luminous X-ray sources, ULX, dot-dashed red line); ii) black hole High-mass X-ray binaries (dashed red line) and iii) accreting neutron stars and pulsars (dotted red line). 
Because of the shallow slope of the X-ray luminosity function of compact X-ray sources in star-forming 
galaxies, X-ray emission of the latter is dominated by the most luminous sources --  stellar mass black 
holes accreting matter from the massive companion star. Adapted from \citet{gilfanov:14}.}
\label{fig:cxrb}
\end{figure*}

In particular, the Cosmic X-ray Background (CXRB) radiation can be considered as
the ultimate inventory of the energy released by the process of accretion onto black holes throughout 
the history of the Universe. Detailed modelling of the CXRB over the years, so-called ``synthesis 
models'' of the CXRB \citep{gilli:07,treister:09}, evolved in parallel with our deeper 
understanding of the physical properties of accreting black holes, and of their cosmological evolution.
Today, deep extragalactic surveys with X-ray focusing telescopes, mainly {\em Chandra} and {\em XMM-Newton}, 
have resolved about $\sim 80-90\%$ of the CXRB. These observations have shown 
unambiguously that a similar fraction of the CXRB emission is produced 
by the emission of supermassive black holes in AGN at cosmological distances (see Figure~\ref{fig:cxrb}).

  The goal of reaching a complete census of
evolving AGN, and thus of the accretion power released by SMBH in the
history of the universe has therefore been intertwined with that of
fully resolving the CXRB into individual sources.  Accurate
determinations of the CXRB intensity and spectral shape, coupled with
the resolution of this radiation into individual sources, allow very
sensitive tests of how the AGN luminosity and obscuration evolve with
redshift.

The most recent CXRB synthesis models
have progressively reduced the uncertainties in the absorbing column
density distribution.  When combined with the observed X-ray
luminosity functions, they provide an almost complete census of the
Compton-thin AGN (i.e., those obscured by columns $N_{\rm
  H}<\sigma_{\rm T}^{-1} \simeq 1.5 \times 10^{24}$ cm$^{-2}$, where
$\sigma_{\rm T}$ is the Thomson cross section).  This class of objects
dominates the counts in the lower energy X-ray energy band, where
almost the entire CXRB radiation has been resolved into individual
sources \cite{worsley:05}.  


Synthesis models of the X-ray background, like the one shown in Figure~\ref{fig:cxrb}
ascribe a substantial fraction of this unresolved emission to heavily
obscured (Compton-thick) AGN. However, because of their faintness even
at hard X-ray energies, their redshift and luminosity distribution is
very hard to determine, and even their absolute contribution to the
overall CXRB sensitively depends on the quite uncertain normalization
of the unresolved emission at hard X-ray energies. Overall the CXRB 
is relatively insensitive to the
precise Compton-thick AGN fraction \citep{akylas:12}. The quest for the
physical characterization of this ``missing'' AGN population, most
likely dominated by Compton thick AGN, represents one of the last
current frontiers of the study of AGN evolution at X-ray wavelengths.

However, Compton Thick AGN still leaves characteristics imprints in the shape of
AGN X-ray spectra, so that the deepest X-ray surveys, along with extensive multi-wavelength
coverage of X-ray survey regions, have allowed the 
identification significant samples of Compton-thick
AGN at moderate to high redshifts \citep{brightman:12,georgantopoulos:13,buchner:14,brightman:14}.

Recently, \citet{buchner:15} developed a novel non-parametric
method for determining the space density of AGN as a
function of accretion luminosity, redshift and hydrogen
column density, building on the X-ray spectral analysis
of \citet{buchner:14}. Applying their Bayesian spectral
analysis technique of a realistic, physically motivated
model to a multi-layered survey determine the luminosity
and level of obscuration in a large sample of X-ray
selected AGN across a wide range of redshifts.

\begin{figure*}
\centering
\includegraphics[width=\textwidth]{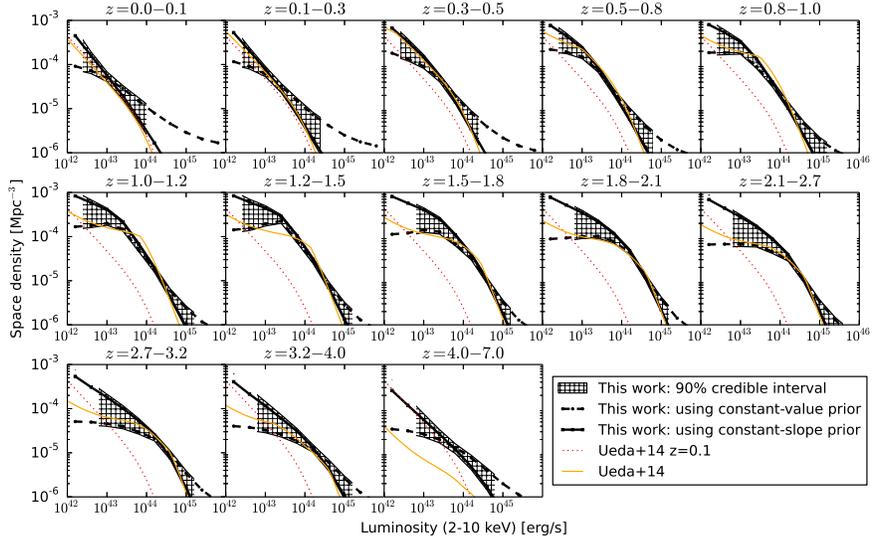}
\caption{Total X-ray luminosity function in the 2-10 keV spectral band. Each panel corresponds to a different redshift interval. Dashed and solid black lines represent the results obtained with two different generic choices of priors for the non-parametric description of
the data; the difference in the reconstructions between the two is therefore an indication of whether
the data or the priors dominate the result. The hatched regions indicate a measure of the uncertainty, using the 10\%-90\% 
quartiles of the posterior samples from both priors together. 
The orange thin solid line shows the reconstruction by \citet{ueda:14}. The dotted red
curve is their local ($z = 0.1$) luminosity function kept constant across all panels for comparison. From \citet{buchner:15}}.
\label{fig:lf_buchner}
\end{figure*}

Figure~\ref{fig:lf_buchner} shows the total (i.e. including Compton-thick objects) X-ray luminosity function (XLF) 
in the 2-10 keV spectral band, together with a comparison with a recent comprehensive study of the
XLF by Ueda et al. (2014). The overall shape of the
luminosity function is a double power-law with a break
or bend at a characteristic luminosity, $L*$, the value
of which increases with redshift. As found in previous
studies, the space density shows a rapid evolution up to
around $z\approx 1$ at all luminosities, being most prominent
at high luminosities due to the positive evolution of $L*$.

\begin{figure*}
\centering
\includegraphics[width=0.8\textwidth]{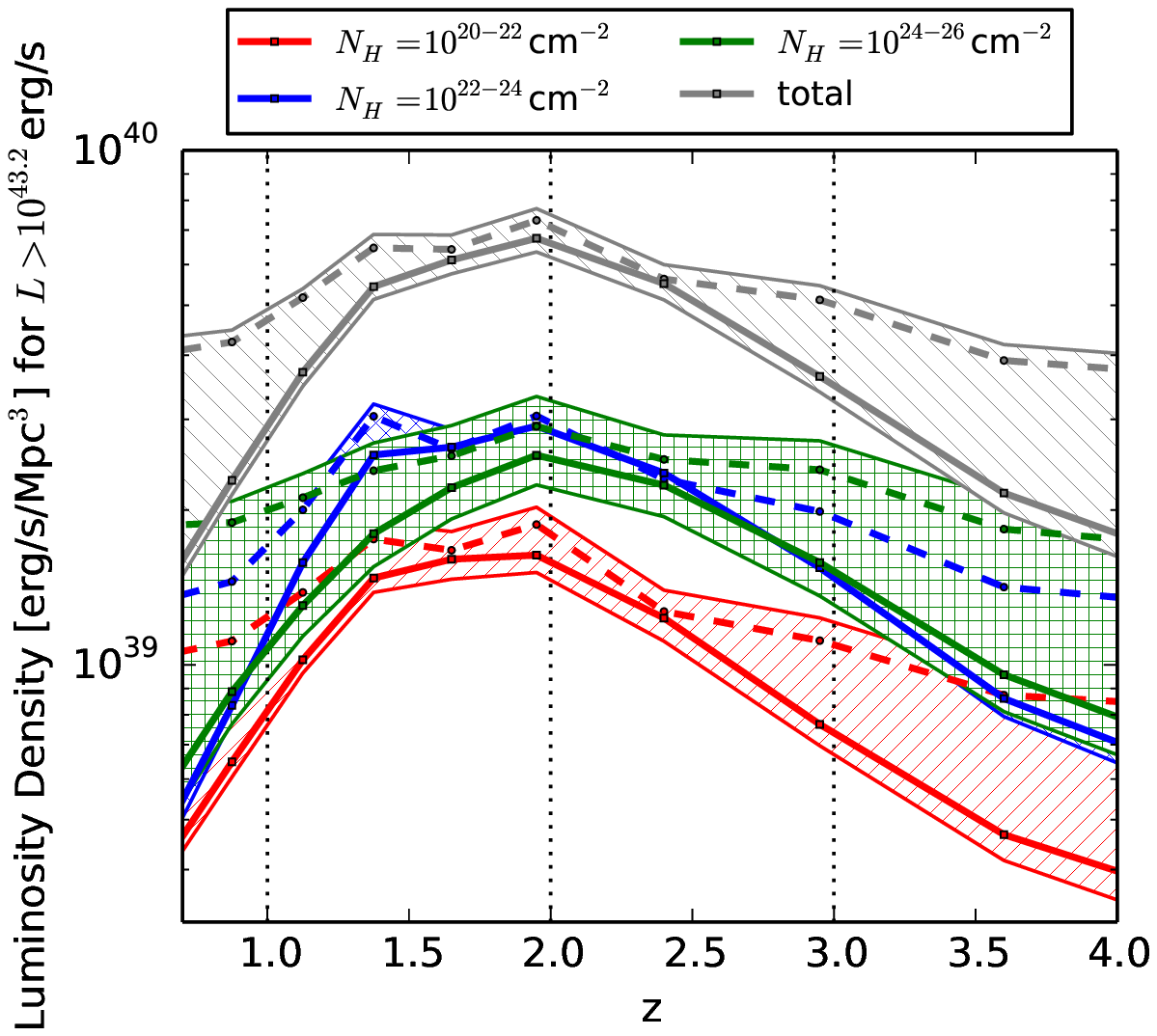}\\
\includegraphics[width=\textwidth]{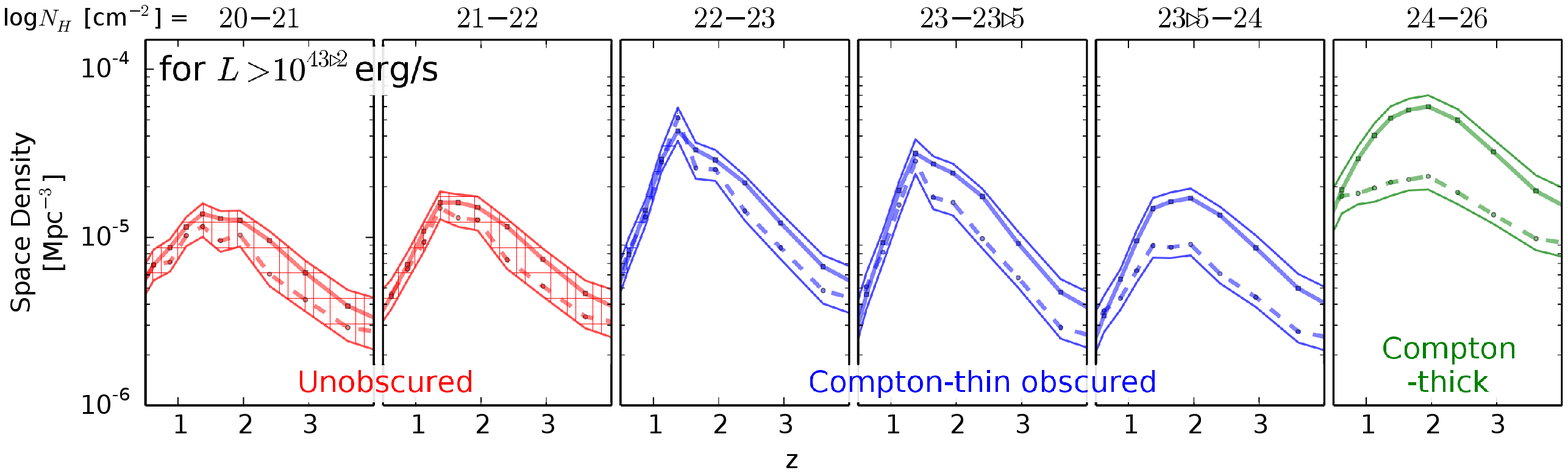}\\
\caption{{\it Top panel:} Evolution of the X-ray luminosity density of AGN
with $L_{\rm X} > 10^{43.2}$ erg/s, for various column densities. The luminosity
output of AGN experiences a rise and fall in density in
the $z = 1 -- 3.5$ range (total as top gray shaded region). The
strongest contribution to the luminosity density is due to obscured,
Compton-thin (blue shaded region) and Compton-thick
AGN (green shaded region), which contribute in equal parts to
the luminosity. The emission from un-obscured AGN
(red shaded region, bottom) is significantly smaller.
{\it Bottom panel:} Redshift evolution of space density of AGN split by the level of obscuration. 
Different panels correspond to different hydrogen
column density interval as indicated at the top. From \citet{buchner:15}.}
\label{fig:space_density_jb}
\end{figure*}

Figure~\ref{fig:space_density_jb} shows the derived luminosity density (in erg/s/Mpc$^{-3}$) for X-ray emitting AGN,
split into un-obscured ones, Compton-thin and -thick sources. \citet{buchner:15} found that about 75\%
of the AGN space density, averaged over redshift, corresponds
to sources with column densities $N_{\rm H} > 10^{22}$ cm$^{-2}$. 
The contribution of obscured AGN to
the accretion density of the Universe over cosmic time is similarly large
($\approx 75$\%). The contribution to the luminosity density by Compton-thick
AGN is $39\pm 6$\%. Crucially, for the first time the uncertainty on what
used to be called ``missing'' AGN population is below 10\%, and the Compton-thick AGN fraction
appears consistent with the requirement of CXRB synthesis models, suggesting we are approaching
a reliable, comprehensive census of accretion onto Supermassive black hole over a large fraction of the
age of the Universe.

\subsection{Bolometric AGN luminosity functions and the history of accretion}
\label{sec:bol_lf}

\begin{figure*}
\centering
\includegraphics[width=\textwidth]{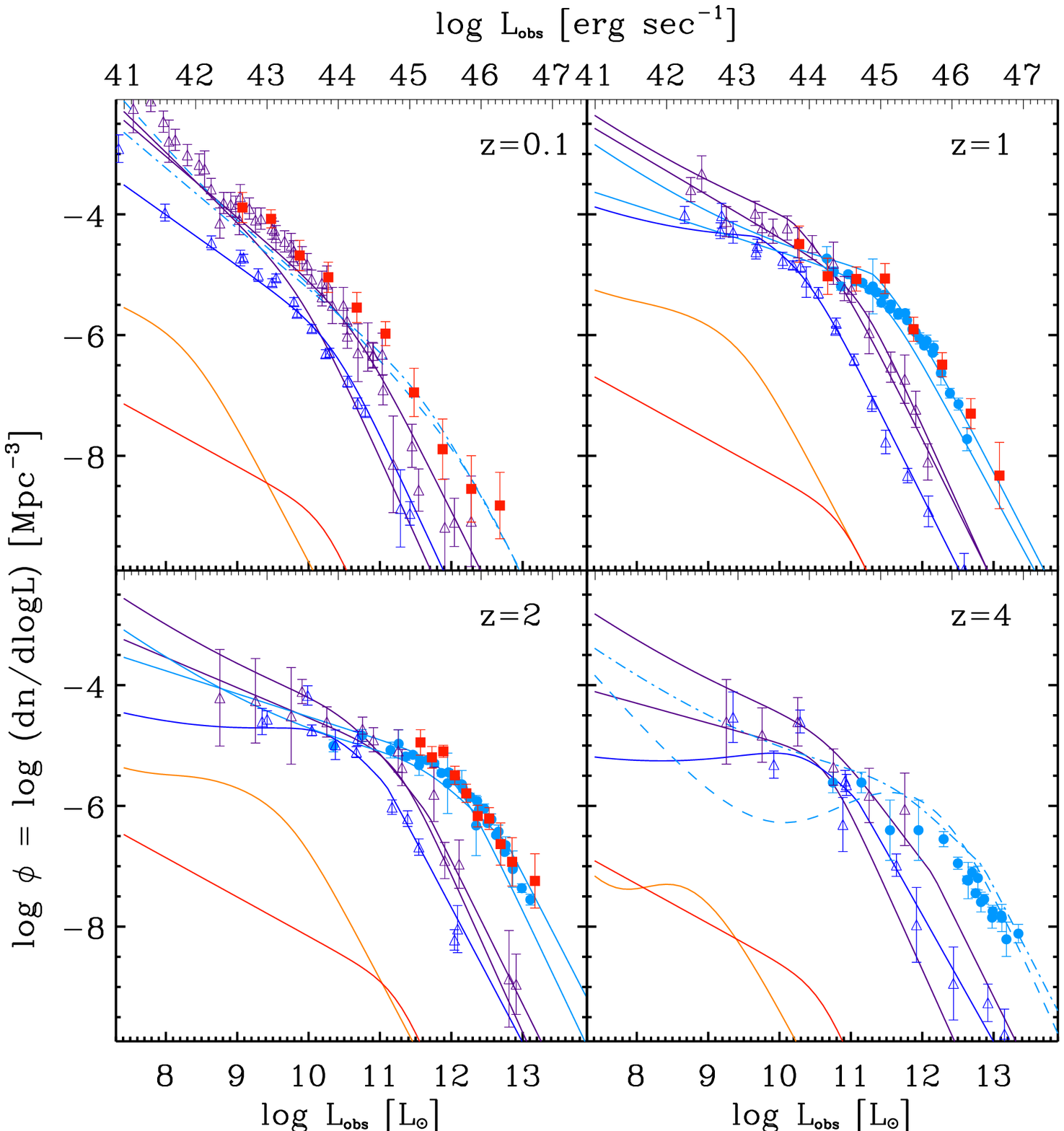}
\caption{A compilation of luminosity functions observed in various
  energy bands.  The logarithm of the number of AGN per unit comoving
  volume and unit logarithm of luminosity is plotted as a function of
  the observed luminosity (in solar units). Observational points for
  IR (15$\mu$m; filled red 
  squares), B-band (filled blue circles), soft- (0.5-2 keV; empty blue
  triangles) and hard-X-rays (2-10 keV; empty purple triangles) are
  shown alongside published analytic fits for each band (solid lines
  in corresponding colors). The best-fit radio luminosity functions of steep- and flat-spectrum
radio sources from Massardi et al. (2010) are also shown for comparison, with orange and red
thick lines, respectively.
   The observed mismatch
  among the various luminosity functions in Fig.~\ref{fig:lf_obs_all}
  is due to a combination of different bolometric corrections and
  incompleteness due to obscuration.  Courtesy of P.~Hopkins}
\label{fig:lf_obs_all}
\end{figure*}

As we have seen in the previous sections, a qualitatively consistent
picture of the main features of AGN evolution is emerging from the
largest surveys of the sky in various energy bands.  Strong (positive)
redshift evolution of the overall number density, as well as some
differential evolution (with more luminous sources being more dominant
at higher redshift) characterize the evolution of AGN (see Figure~\ref{fig:lf_obs_all}).

\begin{figure*}
\centering
\begin{tabular}{cc}
  \includegraphics[width=0.49\textwidth]{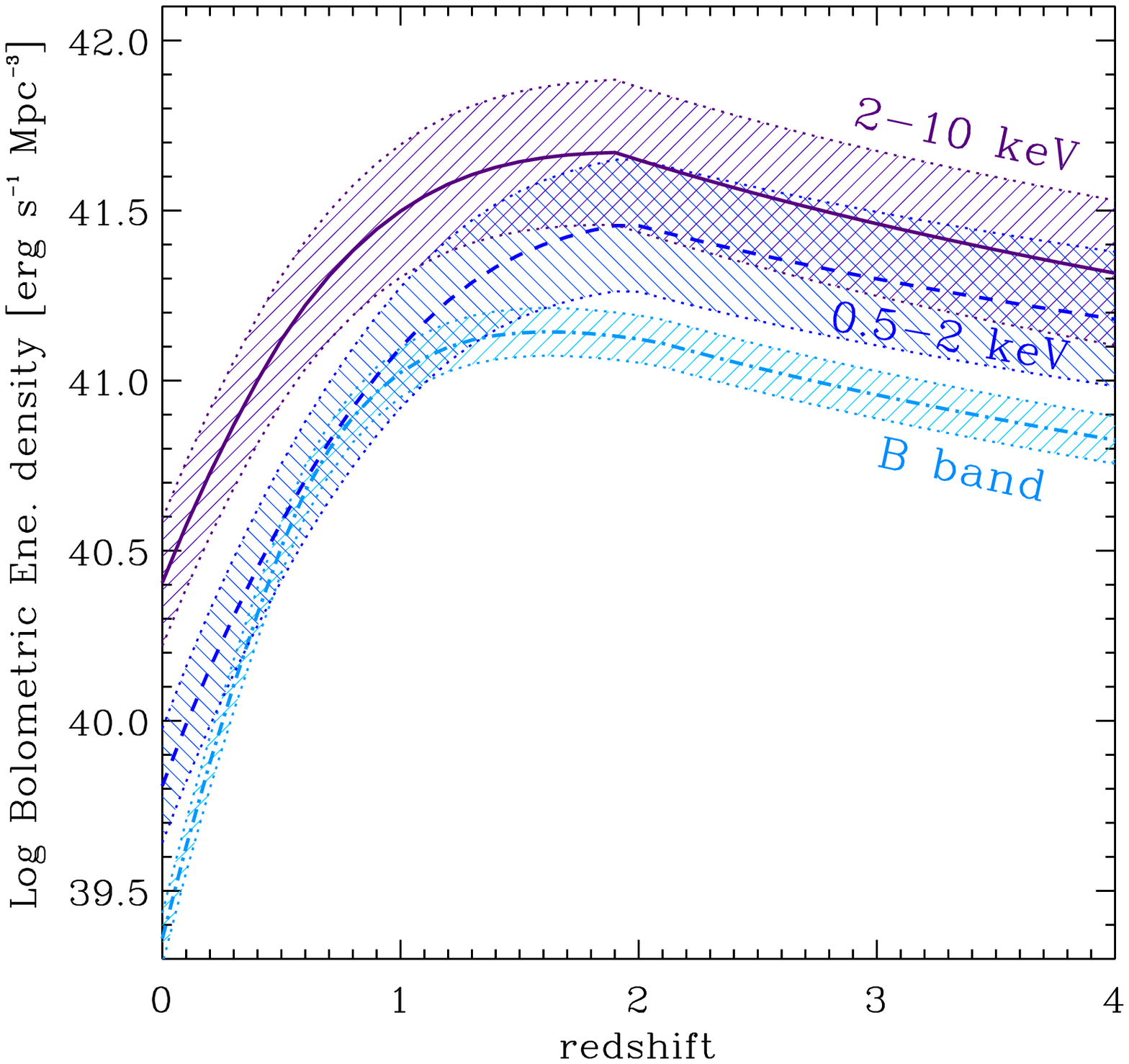} &
  \includegraphics[width=0.49\textwidth]{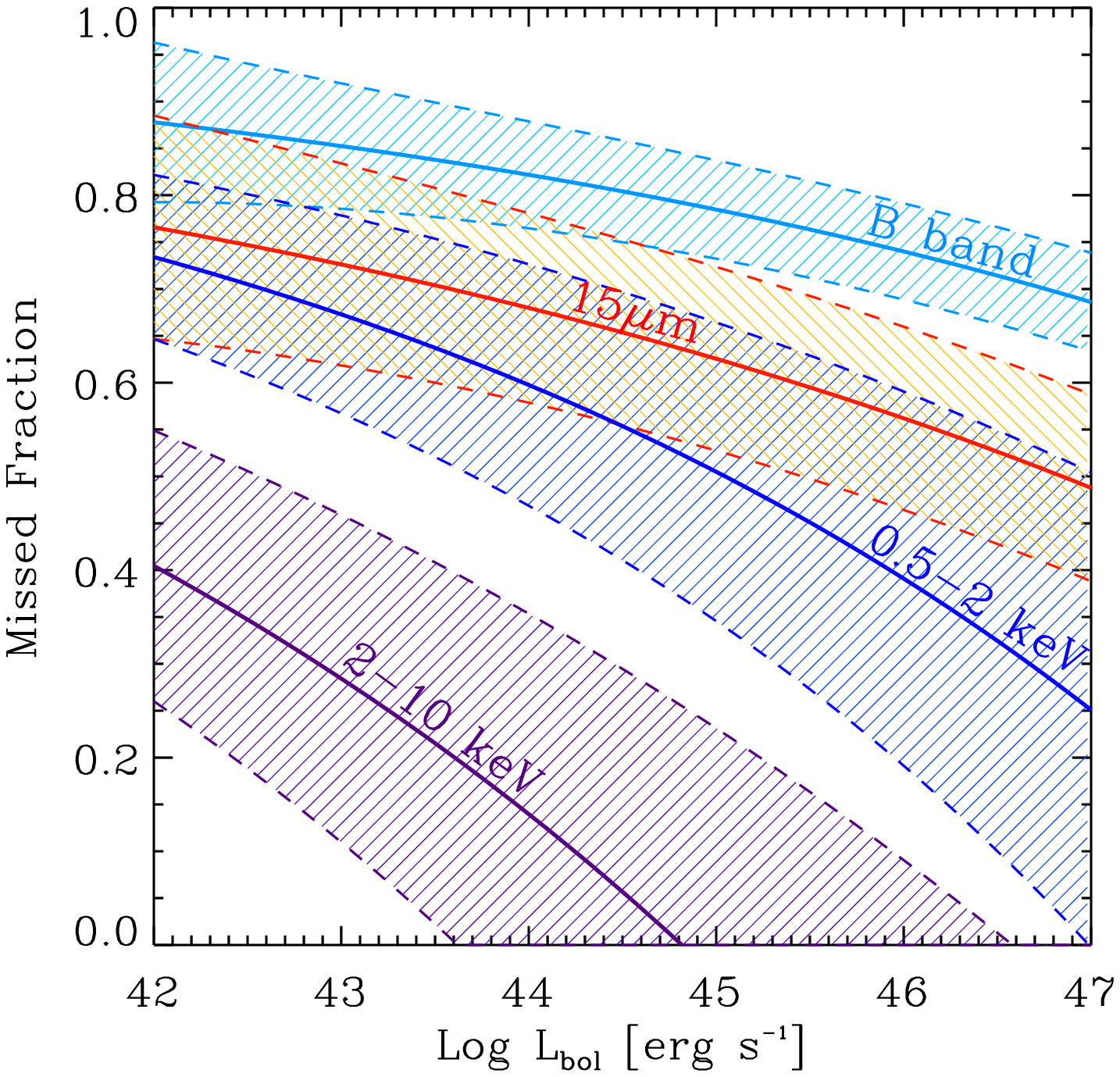} \\
\end{tabular}
\caption{{\it Left:} The redshift evolution of the bolometric energy
  density for AGN selected in different bands.  Bolometric corrections
  from \cite{hopkins:07} have been used, and the shaded areas
  represent the uncertainty coming from the bolometric corrections
  only.  {\it Right:} The fraction of AGN missed by observations in
  any specific band as a function of the intrinsic bolometric
  luminosity of the AGN.  Red, light blue, dark blue and purple shaded
  areas correspond to rest-frame mid-IR (15$\mu$m), UV (B-band), soft
  X-rays (0.5-2 keV) and hard X-rays (2-10 keV), respectively.  The
  uncertainty on the missed fractions depend on the uncertainties of
  the bolometric corrections and on the shape of the observed
  luminosity functions only.}
\label{fig:dens_bol}
\end{figure*}

A thorough and detailed understanding of the AGN SED as a function of
luminosity  could in principle allow us to
compare and cross-correlate the information on the AGN evolution
gathered in different bands. A luminosity dependent bolometric correction is
required in order to match type I (unabsorbed) AGN luminosity
functions obtained by selecting objects in different bands. This is, in a nutshell,
a direct consequence of the observed trend of the relative contribution of optical
and X-ray emission to the overall SED (the $\alpha_{\rm ox}$ parameter) 
as a function of luminosity (see the left panel of Fig.~\ref{fig:coronae_agn}).

Adopting a general form of luminosity-dependent bolometric
correction, and with a relatively simple parametrization of the effect
of the obscuration bias on the observed LF, \cite{hopkins:07} were
able to project the different observed luminosity functions in various
bands into a single bolometric one, $\phi(L_{\rm bol})$.  
As a corollary from such an exercise, 
we can then provide
a simple figure of merit for AGN selection in various bands by
measuring the bolometric energy density associated with AGN selected
in that particular band as a function of redshift. I show this in the
left panel of Figure~\ref{fig:dens_bol} for four specific bands (hard
X-rays, soft X-rays, UV, and mid-IR). From this, it is obvious that
the reduced incidence of absorption in the 2-10 keV band makes the 
hard X-ray surveys recover a higher fraction of the accretion power
generated in the universe than any other method.

While optical QSO surveys miss more than three quarters of all AGN of
any given $L_{\rm bol}$, hard X-ray selection only fails to account
for about one third  of all AGN, the most heavily obscured
(Compton-thick) ones, as shown in the right panel of
Figure~\ref{fig:dens_bol}.  
It is important to note that the
high missed fraction for mid-IR selected AGN is a direct consequence
of the need for (usually optical) AGN identification of the IR
sources, so that optically obscured active nuclei are by and large
missing in the IR AGN luminosity functions considered here.

\section{The Soltan argument: the Efficiency of Accretion}
\label{sec:soltan}
A reliable census of the bolometric energy output
of growing supermassive black holes allows a more direct estimate
of the global rate of mass assembly in AGN, and an interesting
comparison with that of stars in galaxies.  Together with the tighter
constraints on the ``relic'' SMBH mass density in the local universe,
$\rho_{\rm BH,0}$, provided by careful application of the scaling
relations between black hole masses and host spheroids, this enables
meaningful tests of the classical 'Soltan argument' \citep{soltan:82},
according to which the local mass budget of black holes in galactic
nuclei should be accounted for by integrating the overall energy
density released by AGN, with an appropriate mass-to-energy conversion
efficiency.

Many authors have carried out such a calculation, either using the
CXRB as a ``bolometer'' to derive the total energy density released by
the accretion process \citep{fabian:99}, or by considering evolving
AGN luminosity functions \citep{yu:02,marconi:04,merloni:08}.  Despite
some tension among the published results that can be traced back to
the particular choice of AGN LF and/or scaling relation assumed to
derive the local mass density, it is fair to say that this approach
represents a major success of the standard paradigm of accreting black
holes as AGN power-sources, as the radiative efficiencies needed to
explain the relic population are within the range $\approx 0.06 \div
0.40 $, predicted by standard accretion disc theory
\cite{shakura:73}.

In general, we can summarize our current estimate of the (mass-weighted) 
average radiative efficiency, $\langle \epsilon_{\rm rad} \rangle$, together
with all the systematics uncertainties, within one formula, 
relating $\langle \epsilon_{\rm rad} \rangle$ to various sources of systematic errors in
the determination of supermassive black hole mass density. From the integrated bolometric 
luminosity function, we get:

\begin{equation}
\label{eq:soltan}
\frac{\langle \epsilon_{\rm rad} \rangle}{1-\langle \epsilon_{\rm rad} \rangle} \approx 0.075 \left[\xi_0(1-\xi_i-\xi_{\rm CT}+\xi_{\rm lost}) 
\right]^{-1} 
\end{equation}
where $\xi_0=\rho_{\rm BH,z=0}/ 4.2\times 10^5 M_{\odot} {\rm Mpc}^{-3}$ is the local ($z=0$) SMBH
mass density in units of  4.2$\times 10^5 M_{\odot} {\rm Mpc}^{-3}$ \cite{marconi:04}; 
$\xi_i$ is the mass density of black holes at the highest redshift probed by the bolometric
luminosity function, $z \approx 6$, in units of the local one,
 and encapsulate our uncertainty on the process of 
BH formation and seeding in proto-galactic nuclei \cite[see e.g.][]{volonteri:10}; 
$\xi_{\rm CT}$ is the fraction of SMBH mass density (relative to the local one) 
grown in heavily obscured, 
Compton Thick AGN; 
finally, $\xi_{\rm lost}$ is the fraction black hole mass contained in
``wandering'' objects, that 
have been ejected from a galaxy nucleus following, for example, a merging event and the
subsequent production of gravitational wave, the net momentum of which could induce a 
kick capable of ejecting the black hole form the host galaxy.  

\begin{figure*}
\centering
  \includegraphics[width=0.7\textwidth]{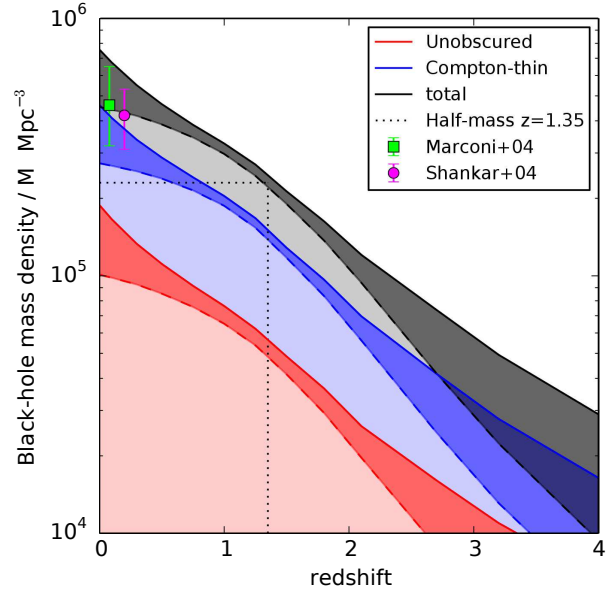} 
\caption{Redshift evolution of the total supermassive black holes mass density. The dark grey band shows the overall evolution, computed
from the observed 2-10 keV luminosity function of Figure~\ref{fig:lf_buchner}, assuming, for simplicity, a constant bolometric 
correction. The red band shows the un-obscured AGN contribution only,
while the blue one is for Compton-thin objects, including un-obscured ones. Colored points with error bars represent estimates of the local SMBH mass density from \citet{marconi:04} and \citet{Shankar:04}. Courtesy of J. Buchner.}
\label{fig:rho_ev}
\end{figure*}

The recent progresses on the tracking of heavily obscured AGN in deep X-ray surveys that we have highlighted in 
section~\ref{sec:xray_surveys} \citep{buchner:15} allows us to estimate the contributions of un-obscured AGN, 
Compton-thin and Compton-thick obscured AGN separately. Figure~\ref{fig:rho_ev} shows an illustrative example 
of such a calculation, where, for the sake of simplicity, the X-ray radiative energy density evolutions of \citet{buchner:15} 
have been used, 
assuming a constant bolometric correction and fixed the radiative efficiency of accretion to 10\%. According to this computation,
the fraction of black hole mass accreted in heavily obscured (CT) phases is $\xi_{\rm CT}\approx$ 35\% (see
Fig.~\ref{fig:space_density_jb}), with $\xi_i<4$\% (at $z=4$).  

In the real evolution of SMBH, all the terms in eq.~(\ref{eq:soltan}) must be linked at some level, as the radiative efficiency of accretion depends on the 
location of the innermost stable circular orbit, thus on the black hole spin, which itself evolves under the effects of both
accretion and BH-BH mergers. Accurate models which keep track of both mass and spin evolution of SMBH could in principle be used, together
with observational constraints from the AGN luminosity functions, to put constraints on those unknowns, providing a direct link 
between the relativistic theory of accretion and structure formation in the Universe \cite[see e.g.][]{volonteri:13,sesana:14}.
Pinning down the uncertainty of the luminosity density of accretion and of the bolometric corrections (properly including its dependence 
on black hole mass, accretion rate, possible redshift) will provide strong indirect bounds on the allowed population of 
seed and ``wandering'' black holes.

\subsection{Quantifying the efficiency of kinetic energy release}

We close this chapter with a brief discussion of the available estimates of the efficiency with which AGN
are able to convert gravitational potential energy of the accreted matter into kinetic energy
of the radio-emitting relativistic jets. This is a fundamental question for our understanding of accretion processes at low
rates, with potentially crucial implication for the physical nature of feedback from AGN.
To carry out this  exercise, however, we need first to measure the total kinetic
power carried by those jets. 

The observed omni-presence of radio cores\footnote{The ``core'' of a
  jet is the brightest innermost region of the jet, where the jet just
  becomes optically thin to synchrotron self absorption, i.e., the
  synchrotron photosphere of the jet.} in low luminosity AGN and the
observed increase in radio loudness of X-ray binaries at low
luminosities can be placed on a solid theoretical footing.  Jets
launch in the innermost regions of accretion flows around black holes,
and at low luminosities, these flows likely become mechanically (i.e.,
advectively) cooled.

Such flows can, to lowest order, be assumed to be scale invariant: a
low luminosity accretion flow around a 10 solar mass black hole,
accreting at a fixed, small fraction of the Eddington accretion rate,
will be a simple, scaled down version of the same flow around a
billion solar mass black hole (with the spatial and temporal scales
shrunk by the mass ratio).  It follows, then, that jet formation in
such a flow should be similarly scale invariant.

This assumption is sufficient to derive a very generic relation
between the radio luminosity emitted by such a scale invariant jet and
the total (kinetic and electromagnetic) power carried down the jet,
{\em independent} of the unknown details of how jets are launched and
collimated \citep{heinz:03}: The synchrotron radio luminosity
$L_{\nu}$ of a self-absorbed jet core depends on the jet power $P_{\rm
  jet}$ through
\begin{equation}
  L_{\rm radio} \propto P_{\rm jet}^{\frac{17+8\alpha}{12}}M^{-\alpha} \sim
  P^{\frac{17}{12}}\label{eq:scaling}
\end{equation}
where $M$ is the mass of the black hole and $\alpha \sim 0$ is the
observable, typically flat radio spectral index of the synchrotron
power-law emitted by the core of the jet.  This relation is a result
of the fact that the synchrotron photosphere (the location where the
jet core radiates most of its energy) moves further out as the size
scale and the pressure and field strength inside the jet increase
(corresponding to an increase in jet power).  As the size of the
photosphere increases, so does the emission.  The details of the
power-law relationship are an expression of the properties of
synchrotron emission.

For a given black hole, the jet power should depend on the accretion
rate as $P_{\rm jet} \propto \dot{M}$ (this assumption is implicit in
the assumed scale invariance).  On the other hand, the emission from
optically thin low luminosity accretion flows itself depends
non-linearly on the accretion rate, roughly as $L_{\rm acc} \propto
\dot{M}^2$, since two body processes like bremsstrahlung and inverse
Compton scattering dominate, which depend on the square of the
density.  Thus, at low accretion rates, $L_{\rm radio} \sim L_{\rm
  bol}^{\frac{17}{24}}$, which implies that black holes should become
more radio loud at lower luminosities
\citep{heinz:03,mhd03,falcke:04}.  It also implies that more
massive black holes should be relatively more radio loud than less
massive ones, at the same {\em relative} accretion rate $\dot{M}/M$.

Equation \ref{eq:scaling} is a relation between the observable core
radio flux and the underlying jet power.  Once calibrated using a
sample of radio sources with known jet powers, it can be used to
estimate the jet power of other sources based on their radio
properties alone (with appropriate provisions to account,
statistically, for differences in Doppler boosting between different
sources).

Nearby radio sources in massive clusters, where the cavities inflated by the relativistic jets can be
used as calorimeters to estimate the total kinetic jet power \citep{allen:06,rafferty:06,merloni:07,cavagnolo:10} 
provide such a sample.  Plotting the core (unresolved) radio power against the jet
power inferred from cavity and shock analysis shows a clear non-linear
relation between the two variables \citep{merloni:07}.  Fitting this
relation provides the required constant of proportionality and is
consistent (within the uncertainties) with the power-law slope of
$17/12$ predicted by eq.~(\ref{eq:scaling})
\begin{equation}
  P_{\rm jet} = P_{0}\left(\frac{L_{\rm core}}{L_{0}}\right)^{\zeta}
  \sim 1.6\times 10^{36}\,{\rm ergs\,s^{-1}} \left(\frac{L_{\rm
        core}}{10^{30}\,{\rm ergs\,s^{-1}}}\right)^{0.81}
  \label{eq:power}
\end{equation}
with an uncertainty in the slope $\zeta$ of 0.11, where $L_{\rm
  radio}=\nu L_{\nu}$ is measured at $\nu=5\,GHz$.

Because this relation was derived for the {\rm cores} of jets, which
display the characteristic flat self-absorbed synchrotron spectrum,
care has to be taken when applying it to a sample of objects: only the
core emission should be taken into account, while extended emission
should be excluded.  Moreover, the jet have relativistic bulk motions on the scales probed by the 
core emission, and the additional effect of relativistic beaming on the shape of the luminosity function
has to be taken into account \citep{merloni:07}.
As discussed in \S\ref{sec:radio_lf}, radio
luminosity functions are separated spectrally into flat and steep
sources, and we can use both samples to limit the contribution of flat
spectrum sources from both ends.

On the other hand, the same sample of radio AGN within clusters of galaxies 
with measured kinetic power can be used to derive the relationship between the
extended, steep radio synchrotron luminosity and the jet power \citep{birzan:08,cavagnolo:10}.
This has the advantage of being an isotropic luminosity indicator (i.e. unaffected by relativistic beaming, as in the 
case of the radio cores), but is also more sensitive to the environment (its density, magnetic field, kinematical state)
the jet impinges upon. Indeed, the most recent analysis reveals a correlation between kinetic power and low-frequency (1.4 GHz) diffuse lobe emission of the form:
\begin{equation}
  P_{\rm jet} \simeq 6.3\times 10^{36}\,{\rm ergs\,s^{-1}} \left(\frac{L_{\rm
        lobe}}{10^{30}\,{\rm ergs\,s^{-1}}}\right)^{0.7}
  \label{eq:power_lobes}
\end{equation}

Given a radio luminosity function $\Phi_{\rm rad}$ (and, in the case of flat spectrum radio cores, an appropriate
correction for relativistic boosting) eq.~\ref{eq:power} and eq.~\ref{eq:power_lobes} can be used
to derive the kinetic luminosity function of jets
\citep{heinz:07,merloni:08}:
\begin{equation}
  \Phi_{\rm kin}(P_{\rm jet}) = \Phi_{\rm
    rad}\left[L_{0}\left(\frac{P_{\rm
          jet}}{P_{0}}\right)^{\frac{1}{\zeta}}\right]
  \frac{1}{\zeta}\frac{L_{0}}{P_{0}}
  \left(\frac{P_{\rm jet}}{P_{0}}\right)^{\frac{1-\zeta}{\zeta}}
\end{equation}
The resulting kinetic luminosity functions (KLF) for the flat spectrum radio
luminosity functions have been presented in \citet{merloni:07}\footnote{Comparison to the steep spectrum
  luminosity function shows that the error in $\Phi_{P}$ from the
  sources missed under the steep spectrum luminosity function is at
  most a factor of two}. They showed that, at the low luminosity end,
these KLF are roughly flat, implying that low luminosity source
contributed a significant fraction of the total power.  These are the
low-luminosity AGN presumably responsible for radio mode feedback, and
they dominate the total jet power output at low redshift.

\begin{figure}[t]
\begin{center}
\resizebox{!}{0.43\textwidth}{\includegraphics{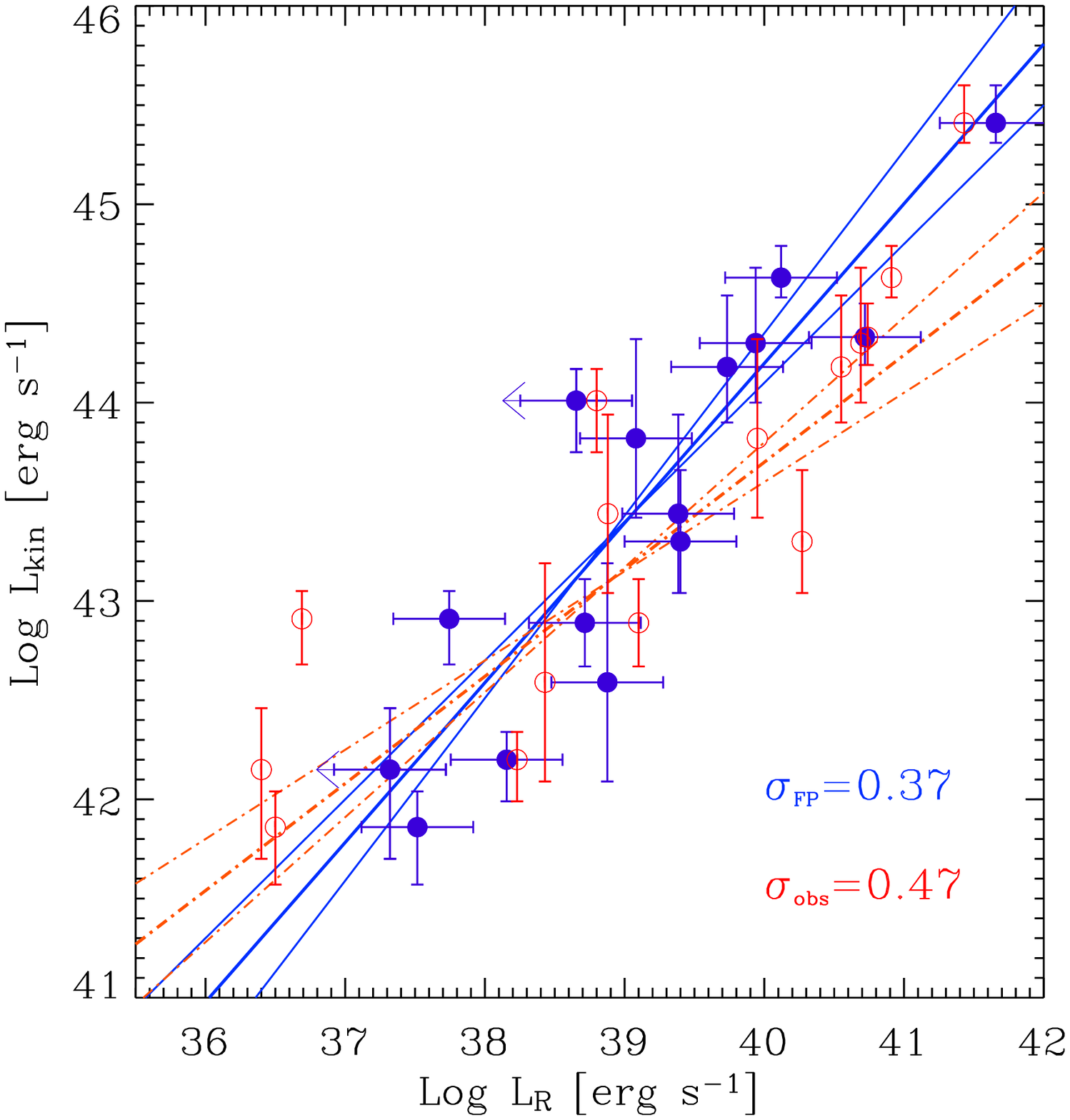}}
\resizebox{!}{0.57\textwidth}{\includegraphics{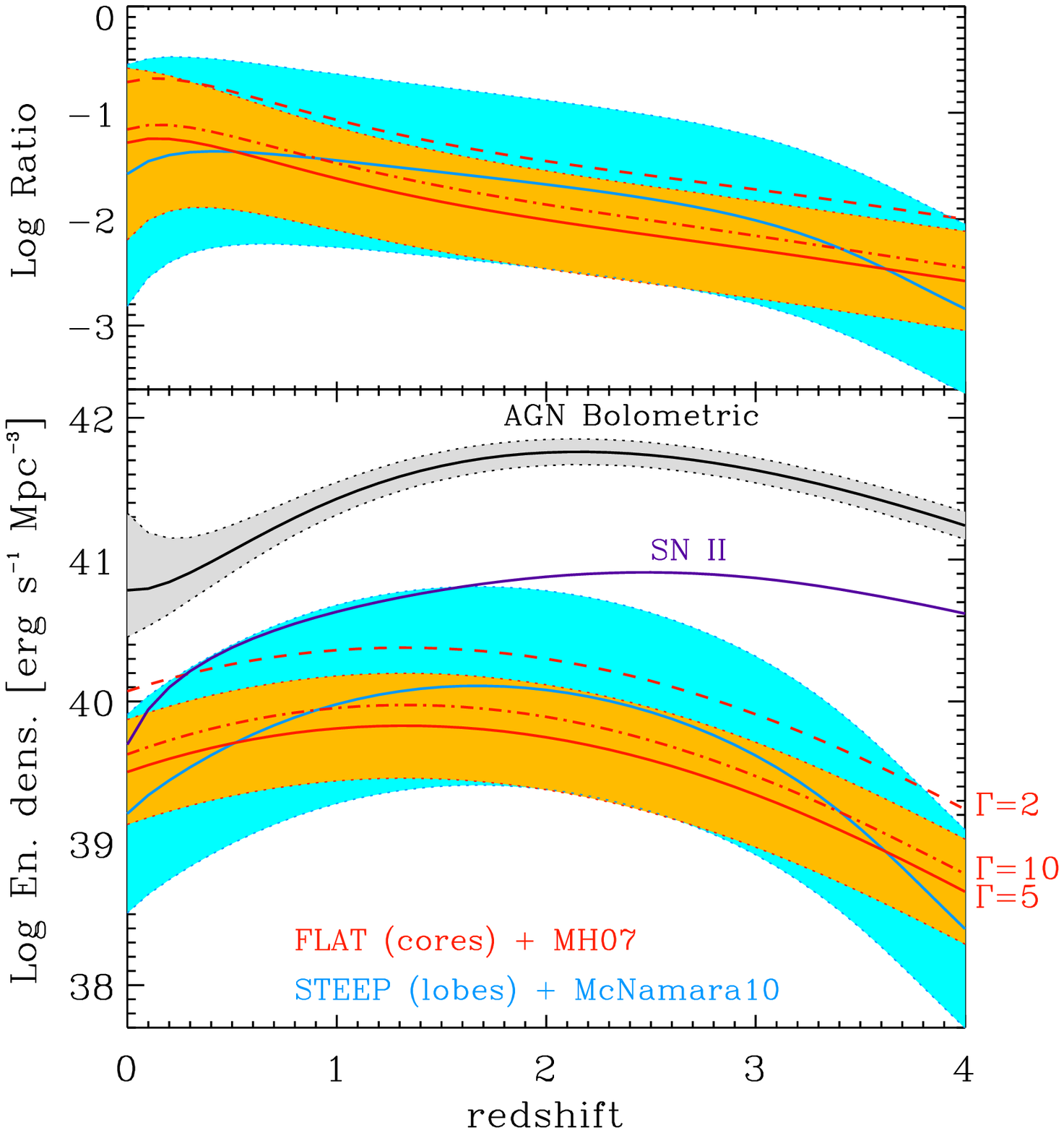}}
\end{center}
\caption{{\it Left}: Red open circles show jet power (measured from
  X-ray cavities) plotted against 5 GHz core radio luminosity (also
  shown in blue solid circles is a Doppler boosting corrected version
  of the same points) along the power-law fit given in
  eq.~(\ref{eq:power}); adopted from \citet{merloni:07}. {\em Right:} The bottom panel
shows the total kinetic energy density (in logarithmic scale) emitted by radio AGN. The red lines
correspond to integrating the flat spectrum radio LF, using eq.(\ref{eq:power}) to obtain the jet power, with
different lines marking the correction due to the assumed average bulk motion Lorentz factor, as labeled. The blue line 
corresponds to integrating the steep spectrum radio LF, using eq.(\ref{eq:power_lobes}) to obtain the jet power. Orange and cyan
bands give an estimate of the uncertainty derived from the observed scatter in those relations. For comparison,
the black line (with grey band) shows the bolometric radiative energy density from AGN, computed from \citet{hopkins:07} bolometric LF, while the purple line is the estimated kinetic power injected into galaxies by core-collapse Supernovae.
The top panel shows the evolution with redshift of the ratio between radiative and kinetic energy density released by growing black holes, separately for flat (red) and steep (blue) spectrum radio sources.
  \label{fig:scaling}}
\end{figure}

One can also integrate the KLF and obtain the total kinetic energy density released by accreting black holes
as radio-emitting relativistic jets. This is shown in the lower panel on the right hand side of Figure~\ref{fig:scaling}, both 
for flat spectrum radio cores (red curves, with different assumptions about the average bulk motion Lorentz factor of the
jets, as labeled), and for the steep spectrum lobes (blue line), each with a nominal uncertainty derived from the scatter about the
relations (\ref{eq:power}) and (\ref{eq:power_lobes}).

Integrating the luminosity function over $P_{\rm jet}$ gives the local
jet power density $\rho_{\rm Pjet}$, which, at redshift zero, is of
the order of $\langle \rho_{\rm Pjet}\rangle \sim 6\times
10^{39}\,{\rm ergs\,s^{-1}\,Mpc^{-3}}$, comparable to the
local power density from supernovae, but will be significantly above
the supernova power in early type galaxies (which harbor massive black
holes prone to accrete in the radio mode but no young stars and thus
no type 2 supernovae).

Finally, integrating $\Phi_{\rm kin}$ over redshift gives the total
kinetic energy density $u_{\rm Pjet}$ released by jets over the
history of the universe, $u_{\rm Pjet} \sim 3\times 10^{57}\,{\rm
  ergs\,Mpc^{-3}}$.  By comparing this to the local black hole mass
density $\rho_{\rm BH}$ we can derive the average conversion
efficiency $\eta_{\rm jet}$ of accreted black hole mass to jet power:
$  \eta_{\rm jet} \equiv u_{\rm Pjet}/\rho_{\rm BH}c^2 \approx 0.2\% -0.5\%$

In other words, about half a percent of the accreted black hole rest
mass energy gets converted to jets, {\em averaged} over the growth
history of the black hole. 

For an average radiative efficiency of about $\langle \epsilon_{\rm rad} \rangle \approx 0.1$, as derived from 
the Soltan argument and discussed in the previous section, the mean kinetic-to-radiative power ratio of AGN
is of the order of a few per cent, but possibly redshift dependent (see the top right panel of Fig.~\ref{fig:scaling}).
However, since most black hole mass was accreted during the quasar epoch, when
black holes were mostly radio quiet, about 90\% of the mass of a given
black hole was accreted at zero efficiency (assuming that only 10\% of
quasars are radio loud).  Thus, the average jet production efficiency
during radio loud accretion must be at least a factor of 10 higher,
about 2\%-5\%, comparable to the {\em radiative} efficiency of
quasars.

\begin{acknowledgement}
I thank the editor, Francesco Haardt and 
all the organizers and students of the SIGRAV PhD school for the generous invitation to write this review and for 
a great and stimulating environment, during which this 
work was conceived. I also thank the editors for the patience and endurance they have demonstrated in coping with my 
delays. This work could not have been possible without the contribution from many collaborators, in particular
Angela Bongiorno, Johannes Buchner, Marat Gilfanov, Sebastian Heinz and Marta Volonteri, who provided ideas, suggestions and
material for this work. I also thank Philip Best, Murray Brightman, Marcella Brusa, Andrea Comastri, 
Ivan Delvecchio, Chris Done, Rob Fender, Antonis Georgakakis, Gabriele Ghisellini, Roberto Gilli, Phil Hopkins, Beta Lusso,
Kirpal Nandra, Paolo Padovani and Mara Salvato for the 
useful discussions.
\end{acknowledgement}
\bibliographystyle{aa} \bibliography{como}
\end{document}